\documentclass[prl,twocolumn,aps,showpacs]{revtex4}

\usepackage{epsfig}
\usepackage{graphicx}
\usepackage{color}

\newcommand{\gapprox}{\stackrel{>}{\sim}}

\begin{document}

\title{Dispersions, weights, and widths of the single-particle spectral function \\ in the normal phase of a Fermi gas}

\author{F. Palestini, A. Perali, P. Pieri, and G. C. Strinati}

\affiliation{Dipartimento di Fisica, Universit\`{a} di Camerino, 62032 Camerino, Italy}


\begin{abstract}
The dispersions, weights, and widths of the peaks of the single-particle spectral function in the presence of pair correlations, for a Fermi gas with \emph{either} attractive \emph{or} repulsive short-range inter-particle interaction, are determined in the normal phase over a wide range of wave vectors, with a twofold purpose.
The first one is to determine how these dispersions identify both an energy scale known as the pseudo-gap near the Fermi wave vector, as well as an additional energy scale related to the contact $C$ at large wave vectors.
The second one is to differentiate the behaviors of the repulsive gas from the attractive one in terms of crossing versus avoided crossing of the dispersions near the Fermi wave vector.
An analogy will also be drawn between the occurrence of the pseudo-gap physics in a Fermi gas subject to pair fluctuations and the persistence of local spin waves in the normal phase of magnetic materials.
\end{abstract}

\maketitle

\section{I. Introduction} 

Local order of short-range nature in the normal phase of an ultra-cold Fermi gas above the superfluid temperature $T_{c}$ has recently been the subject of intense interest, owing to experimental and theoretical advances which have hinged on this local order from different perspectives.

The experimental interest \cite{Jin-2008,JILA-Cam-2010} has mostly focused on the issue of the \emph{pseudo-gap} $\Delta_{\mathrm{pg}}$, which is a \emph{low-energy scale} that in these systems evolves in temperature with continuity out of the pairing gap present in the broken-symmetry (superfluid) phase \cite{Levin-2005}.
On physical grounds, this continuous evolution is due to the persistence of ``medium-range'' pair correlations, which are the remnant above $T_{c}$ of the long-range order below $T_{c}$.

The theoretical interest has been prompted, on the other hand, by the introduction of a number of universal relations due to Tan \cite{Tan-2008-I,Tan-2008-II}, which  are due to the inter-particle interaction being of the contact type and affect several physical quantities. 
These universal relations all depend on a coupling- and temperature-dependent quantity named the \emph{contact} $C$, which can in turn be conveniently expressed in terms of a \emph{high-energy scale} $\Delta_{\infty}$ \cite{PPS-2009}.                              
The fact that $C$ specifies, in particular, the strength of ``short-range'' pair correlations between opposite spins implies that, in ultimate analysis, the high-energy scale $\Delta_{\infty}$ associated with $C$ and the low-energy scale $\Delta_{\mathrm{pg}}$ associated with the pseudo-gap both originate from the same kind of pair correlations, which remain active above $T_{c}$ even in the absence of long-range order.

In this paper, we aim at organizing these two energy scales into a single wave-vector dependent function $\Delta(k)$, of which $\Delta_{\mathrm{pg}}$ represents the value about the Fermi wave vector $k_{F}$ and $\Delta_{\infty}$ its behavior for $k$ much larger than $k_{F}$, corresponding to medium- and short-range pair correlations, in the order.
In practice, from the numerical calculations it is meaningful to determine the values of $\Delta(k)$ just in these two intervals, namely,
for $k \approx k_{F}$ (obtaining $\Delta_{\mathrm{pg}}$) and $k \gg k_{F}$ (obtaining $\Delta_{\infty}$).
This wave-vector dependence arises even though the inter-particle interaction is of the contact type, which at the mean-field level below $T_{c}$ would instead give rise to a wave-vector independent gap.

To this end, we shall analyze in detail the \emph{dispersions\/} of the peaks of the single-particle spectral function for various couplings across the BCS-BEC crossover and temperatures above $T_{c}$, and show how they can rather accurately be represented by BCS-like dispersions with a characteristic ``back-bending'' for the occupied states \cite{Jin-2008,JILA-Cam-2010}.
These dispersions will be obtained within the t-matrix approximation for an attractive inter-particle interaction following the approach of Ref.\cite{PPSC-2002}, which was recently applied to account for the experimental data on ultra-cold Fermi gases \cite{Cam-JILA-2011}.
In addition, we will show that the \emph{weights\/} of the two peaks of the single-particle spectral function can also be described by BCS-like expressions.
Determining these weights will also be useful to obtain the asymptotic value of $\Delta(k)$ for large $k$, where tracing the dispersions may become ill-defined owing to the strong broadening of the large-$k$ structure of the single-particle spectral function at negative frequencies.

The importance of determining the weights (besides than merely focusing on the existence of the pseudo-gap) is in line with the 
emphasis that was given from the early days of the BCS theory of superconductivity to the role of the ``coherence factors''.
Their presence, in fact, made the BCS theory soon accepted as the correct one, because it was then possible to account for the counter-intuitive outcomes of different experiments that could otherwise not be understood only on the basis of the occurrence of a gap in the single-particle spectrum 
\cite{Schrieffer-1964,BCS-50-years}.

The crossed check, between the dispersions and weights of the two branches of the single-particle spectral function, will therefore represent a fingerprint of the survival in the normal phase of typical BCS-like features due to strong pairing fluctuations.
Differences, however, between the broken-symmetry phase below $T_{c}$ and the pseudo-gap phase above $T_{c}$ will mostly appear in the \emph{widths} of the peaks of the single-particle spectral function, which are much broader above than below $T_{c}$ as expected in the absence of a truly long-range order.

In this context, we shall also resume an argument that was raised in Ref.\cite{Randeria-2010}, according to which the occurrence of the above mentioned back-bending  
for $k \gg k_{F}$ should not reflect \emph{per se} the presence of a pseudo-gap for $k$ about $k_{F}$.
This is because the structure at large $k$, which is related to the contact $C$, can be found even in a normal gas with repulsive inter-particle interaction.

Accordingly, we shall argue that the main differences, between the features of the single-particle spectral function for a Fermi gas with repulsive or attractive interaction  in the normal phase, appear actually for $k$ about $k_{F}$.
Specifically, an \emph{avoided crossing} results in the dispersions of the two peaks of the single-particle spectral function in the attractive case, while a \emph{crossing} occurs in the repulsive case.
In the attractive case, the energy spread of the avoided crossing is directly related to the pseudo-gap energy scale $\Delta_{\mathrm{pg}}$.
On physical grounds, this difference between avoided crossing and crossing is a consequence of particle-hole mixing, which survives at a local level in the attractive case when passing from below to above $T_{c}$ but is absent in the repulsive case.

 The overall purpose here, therefore, is not to establish specific criteria for the existence of a pseudo-gap phase in a Fermi gas with attractive interaction.
Rather, we shall be interested in framing the total amount of information, which can be extracted from the single-particle spectral function of an interacting Fermi gas subject to pairing fluctuations above $T_{c}$, into a unified picture where analogies and differences with respect to a simple BCS-like description below 
$T_{c}$ can be emphasized.
 
It is, nevertheless, relevant to provide at this point an (albeit concise) overview of the major relevant work done previously by several groups on the issue of the pseudo-gap, also to recall how this concept had developed in the context of a Fermi gas with attractive interaction.
The prediction for the existence of a pseudo-gap in the normal phase of strongly interacting ultra-cold Fermions was introduced in Ref.\cite{Levin-2004}
within a two-channel fermion-boson model and in Ref.\cite{PPPS-2004} within a single-channel fermion model, before the observation of superfluidity in these gases.
These works were, in turn, based on earlier studies which applied the physics of the BCS-BEC crossover to the high-temperature cuprate superconductors.
In that context, initial work interpreted the normal state of a superfluid in the crossover regime between BCS and BEC as a phase of uncorrelated 
pairs \cite{Micnas-1990} or as a spin-gap phase \cite{Randeria-1992}.
Later, it was shown that this phase reflects a normal phase pseudo-gap, which displays peculiar features in the fermionic spectral function that reflect the presence of a pairing gap in the superfluid phase \cite{Levin-1997}\cite{PPSC-2002}.
Extensive theoretical work on the pseudo-gap issue for a Fermi gas with attractive interaction was reported more recently in 
Refs.\cite{Bulgac-2009,Ohashi-2009,Levin-2010,Sheehy-2010,Ohashi-2010,Bulgac-2011}.

Finally, in the present paper a similarity will be highlighted  between the pseudo-gap physics resulting from pairing fluctuations above $T_{c}$ and the persistence of spin waves over limited spatial regions in the normal phase of ferromagnetic (or anti-ferromagnetic) materials.
Besides being of heuristic value for envisaging the local order associated with the pseudo-gap, this analogy evidences how the current debate about the occurrence of a pseudo-gap in an ultra-cold Fermi gas retraces a similar debate that went on for some time about the persistence of spin waves in magnetic materials.

The paper is organized as follows.
In Section II, the dispersions, weights, and widths of the peaks in the single-particle spectral function for the attractive case are studied in detail, to determine how the energy scale associated with the pseudo-gap about $k_{F}$ evolves for large $k$ toward the energy scale associated with the contact $C$.
In Section III, the single-particle spectral function for the repulsive case is contrasted with that for the attractive case, to bring out the issue of the crossing vs avoided-crossing of the dispersion relations about $k_{F}$ which clearly differentiates between the two cases.
In Section IV, an analogy is drawn between the pseudo-gap physics and the persistence of spin waves in magnetic materials, and a suggestion is made for an additional experimental evidence for the occurrence of a pseudo-gap.
The Appendix gives analytic details about the treatment of pair fluctuations in the repulsive case to obtain the single-particle spectral function over a wide range of $k$.

\section{II. The attractive case: Pseudo-gap vs contact} 

In this Section, we consider a homogeneous Fermi gas with an attractive interaction $v_{0} \delta(\mathbf{r} - \mathbf{r'})$ of short-range between opposite spin atoms with equal populations, whose strength $v_{0}$ can be eliminated in favor of the scattering length $a_{F}$ via the relation:
\begin{equation}
\frac{m}{4 \, \pi \, a_{F}} \, = \,  \frac{1}{v_{0}} \, + \, \int^{k_{0}} \! \frac{d\mathbf{k}}{(2 \pi)^{3}} \,\,
 \frac{m}{\mathbf{k}^{2}}    \,\, .                                                                                                                   \label{regularization}
\end{equation}

\noindent
Here, $m$ is the particle mass, $\mathbf{k}$ a wave vector, and $k_{0}$ a wave-vector cutoff which can be let $\rightarrow \infty$ while $v_{0} \rightarrow 0$ in order to keep $a_{F}$ at a desired value (we set $\hbar = 1$ throughout).

Since $v_{0} < 0$, $a_{F}$ can be positive as well as negative, and the dimensionless interaction parameter $(k_{F} a_{F})^{-1}$ ranges from 
$(k_{F} a_{F})^{-1} \lesssim -1$ in the weak-coupling (BCS) regime, to $ (k_{F} a_{F})^{-1}\gtrsim +1$ in the strong-coupling (BEC) regime, across the unitary limit where $|a_{F}|$ diverges and $(k_{F}\, a_{F})^{-1}=0$.
In practice, the BCS-BEC crossover region of most interest is limited to the interval $-1 \lesssim (k_{F}\, a_{F})^{-1} \lesssim +1$. 

In the superfluid phase well below $T_{c}$ a description of the BCS-BEC crossover results already at the mean-field level, while in the normal phase above $T_{c}$ inclusion of pairing fluctuations is required to get physically meaningful results.
Pairing fluctuations, in particular, turn the characteristic BCS mean-field energy gap below $T_{c}$ into a pseudo-gap above $T_{c}$, as discussed next.

\vspace{0.05cm}
\begin{center}
{\bf A. Mean-field description below $T_{c}$}
\end{center}
\vspace{0.05cm}

The simplest description of the BCS-BEC crossover results within mean field for temperatures $T$ below $T_{c}$, by supplementing
the equation for the BCS gap $\Delta$ 
\begin{equation}
\int \! \frac{d\mathbf{k}}{(2 \pi)^{3}} \, \left( \frac{ 1 - 2 f(E_{\mathbf{k}}) }{2 E_{\mathbf{k}}} - \frac{m}{\mathbf{k}^{2}}  \right)  
\, = \, - \, \frac{m}{4 \, \pi \, a_{F}}                                                                                                         \label{BCS-gap-equation}
\end{equation}

\noindent
with the density equation 
\begin{eqnarray}
n & = & \int \! \frac{d\mathbf{k}}{(2 \pi)^{3}} \left[ f(E_{\mathbf{k}}) \left(1 + \frac{\xi_{\mathbf{k}}}{E_{\mathbf{k}}} \right) \right.         \nonumber  \\
& + & \left. \left(1 - f(E_{\mathbf{k}}) \right) \left(1 - \frac{\xi_{\mathbf{k}}}{E_{\mathbf{k}}} \right)   \right] \,\, .                                       \label{BCS-density-equation}
\end{eqnarray}

\noindent
Here, $E_{\mathbf{k}} =\sqrt{\xi_{\mathbf{k}}^{2} + \Delta^{2}}$ with $\xi_{\mathbf{k}}=\mathbf{k}^{2}/(2m) - \mu$ and
$f(E) = (e^{E/(k_{B}T)} + 1)^{-1}$ is the Fermi function
($\mu$ being the fermionic chemical potential and $k_{B}$ the Boltzmann constant).
Note that the mean-field gap $\Delta$ does not depend on $k = |\mathbf{k}|$ owing to the short-range nature of the inter-particle interaction.

\begin{figure}[t]
\begin{center}
\includegraphics[angle=0,width=8.5cm]{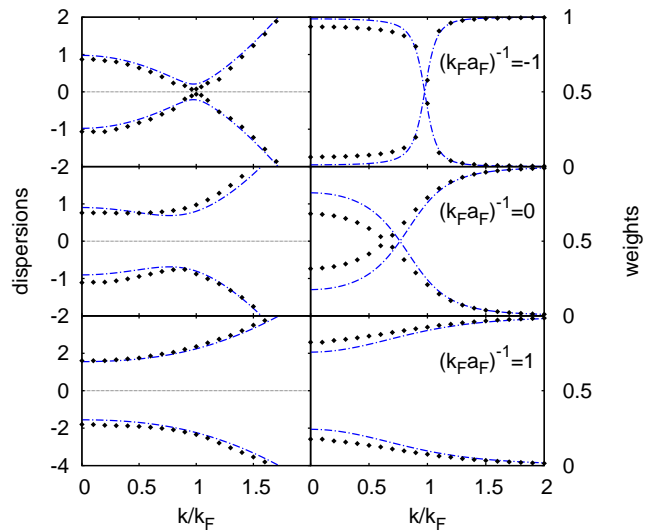}
\caption{Dispersion relations (left panels) and corresponding weights (right panels) for three different couplings, as obtained at $T=0$ within mean field (dashed-dotted lines) and at $T_{c}$ with the inclusion of pairing fluctuations (diamonds).
Energies are in units of $E_{F}$.}
\label{fig-II-1}
\end{center}
\end{figure}

When one looks at the structures of the single-particle spectral function $A(k,\omega)$ within the BCS approximation, at a given $k$ two sharp peaks appear centered at the frequency values
\begin{equation}
\omega = \pm \, E_{\mathbf{k}}                                                                            \label{BCS-dispersion}
\end{equation}

\noindent
with weights $(1 \pm \xi_{\mathbf{k}}/E_{\mathbf{k}})/2$, respectively  \cite{FW}.

The dispersion relations (\ref{BCS-dispersion}) and the corresponding weights are shown in Fig.~\ref{fig-II-1} for three characteristic couplings across the BCS-BEC crossover at zero temperature (dashed-dotted lines).
Here, $E_{F} = k_{F}^{2}/(2m)$ is the Fermi energy with $k_{F} = (3 \pi^{2} n)^{1/3}$.

\vspace{0.05cm}
\begin{center}
{\bf B. Pairing fluctuations above $T_{c}$}
\end{center}
\vspace{0.05cm}

The above picture gets somewhat modified when pairing fluctuations beyond mean field are considered below $T_{c}$ \cite{PPS-2004}.
It is, however, \emph{above} $T_{c}$ that inclusion of pairing fluctuations alters mostly the behavior of $A(k,\omega)$ from its trivial mean-field description with $\Delta = 0$, whereby only a single sharp peak of unit weight survives consistently with a Fermi-liquid picture \cite{Schrieffer-1964}.

In the context of the BCS-BEC crossover, a non-trivial behavior of the spectral function (at and) above $T_{c}$ results when including pairing fluctuations within the t-matrix approximation.
It is still possible to identify two peaks in $A(k,\omega)$ for given $k$ over an extended range of coupling and temperature, by locating their positions and determining their weights and widths. 
It is found that the positions of these peaks can be rather well represented by a BCS-like dispersion of the form (\ref{BCS-dispersion}), provided the mean-field $\Delta$ is replaced by a ``pseudo-gap'' value $\Delta_{\mathrm{pg}}$ that remains finite above $T_{c}$.  
This finding was explicitly demonstrated in Ref.\cite{PPSC-2002} only for the dispersion relations not too far from $k_{F}$ and on the BCS side of the crossover. 

As an example, we report in Fig.~\ref{fig-II-1} the \emph{dispersion relations} and \emph{weights} of the two peaks of $A(k,\omega)$ at $T_{c}$  for three couplings across the BCS-BEC crossover, obtained according to the t-matrix approximation (diamonds) \cite{PPSC-2002}.
In all cases, the similarity with the corresponding values obtained for these quantities within mean field at $T = 0$ (dashed-dotted lines) appears striking.

\begin{figure}[h]
\begin{center}
\includegraphics[angle=0,width=7.0cm]{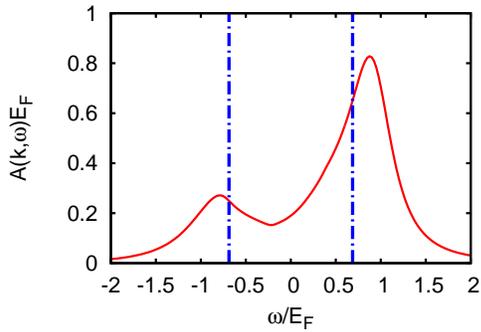}
\caption{Single-particle spectral function $A(k,\omega)$ versus $\omega$ at unitarity, within mean field at $T=0$ (dashed-dotted lines) and with the inclusion of pairing fluctuations at $T_{c}$ (full line), for the wave vector where the maximum of the lower dispersion relation occurs in both cases (see text).}
\label{fig-II-2}
\end{center}
\end{figure}

Marked differences appear instead for the \emph{widths} of the peaks of $A(k,\omega)$, when passing from the mean-field description below $T_{c}$ where they are delta-like, to the t-matrix description (at and) above $T_{c}$ where they are broad and overlapping.
This is shown in Fig.~\ref{fig-II-2}, where $A(k,\omega)$ is plotted versus $\omega$ at unitarity for the wave vector where the maximum of the lower dispersion relation occurs (that is, $0.76 k_{F}$ within mean field at $T=0$ and $0.91 k_{F}$ with the inclusion of pairing fluctuations at $T_{c}$).
This picture also evidences how a real gap at $T=0$ transforms into a pseudo-gap at $T_{c}$, through a partial filling of the spectral function in the region between the two peaks.
In the present paper we shall dwell extensively on this and related ideas.

The t-matrix approximation that we adopt in this paper to obtain $A(k,\omega)$ above $T_{c}$ corresponds to the following choice of the fermionic self-energy \cite{PPSC-2002}:
\begin{eqnarray}
\Sigma(\mathbf{k},\omega_{n}) = & - & \int \! \frac{d\mathbf{q}}{(2 \pi)^{3}} \, k_{B} T  \,
                                                   \sum_{\nu} \, \Gamma_{0}(\mathbf{q},\Omega_{\nu})                             \nonumber  \\
                                               & \times &     G_{0}(\mathbf{q - k},\Omega_{\nu} - \omega_{n})             \label{Matsubara-self-energy}
\end{eqnarray}

\noindent
where $\omega_{n} = (2 n + 1) \pi k_{B} T$ ($n$ integer) and $\Omega_{\nu} = 2 \nu  \pi k_{B} T$ ($\nu$ integer) are
fermionic and bosonic Matsubara frequencies, in the order, $G_{0}(\mathbf{k},\omega_{n}) = (i \omega_{n} - \xi_{\mathbf{k}})^{-1}$ is the bare fermionic single-particle Green's function, and $\Gamma_{0}(\mathbf{q},\Omega_{\nu})$ is the particle-particle ladder given by
\begin{eqnarray}
- \, \Gamma_{0}(\mathbf{q},\Omega_{\nu})^{-1} & = &  \int \! \frac{d\mathbf{k}}{(2 \pi)^{3}} \,  \left[  k_{B} T  \, \sum_{n} \, G_{0}(\mathbf{k},\omega_{n})  \right.                                                                                                                      \label{particle-particle-ladder}  \\
& \times &  \left. G_{0}(\mathbf{q - k},\Omega_{\nu} - \omega_{n}) \, - \, \frac{m}{ \mathbf{k}^{2} }  \right] \, + \, \frac{m}{4 \pi a_{F}} \,\, .   \nonumber                                                                    
\end{eqnarray}

\noindent
The single-particle spectral function is then obtained through analytic continuation $i \omega_{n} \rightarrow \omega + i \eta$ to the real frequency axis ($\eta = 0^{+}$):
\begin{equation}
A(\mathbf{k},\omega) = - \, \frac{1}{\pi} \,
\frac{\mathrm{Im}\Sigma(\mathbf{k},\omega)}{\left[ \omega - \xi_{\mathbf{k}} - \mathrm{Re}\Sigma(\mathbf{k},\omega) \right]^{2}
+ \left[ \mathrm{Im}\Sigma(\mathbf{k},\omega) \right]^{2}} \,\, .                                       \label{single-particle-spectral-function}
\end{equation}

\noindent
The shape of $A(k,\omega)$ versus $\omega$ thus depends crucially on the interplay between 
$\mathrm{Re}\Sigma(\mathbf{k},\omega)$ and $\mathrm{Im}\Sigma(\mathbf{k},\omega)$ for the chosen value of
$k = |\mathbf{k}|$.

A derived quantity of interest is the single-particle \emph{density of states}:
\begin{equation}
N(\omega) \, = \, \int \! \frac{d\mathbf{k}}{(2 \pi)^{3}} \, A(\mathbf{k},\omega)     \,\, .           \label{DOS}
\end{equation}

\noindent
The averaging that this definition introduces on $A(\mathbf{k},\omega)$ over an extended range of $\mathbf{k}$ can be of support to the presence of a pseudo-gap, in cases when the two peaks of $A(\mathbf{k},\omega)$ strongly overlap just in the range of $\mathbf{k}$ where the two branches of the dispersion come close to each other (cf. Fig.~\ref{fig-II-1}).
In these cases, in fact, a strict definition of the pseudo-gap as a depression of the spectral weight just in this range of $\mathbf{k}$ would lead one to conclude that pseudo-gap phenomena were absent in the single-particle excitations, while they still appear clearly over a more extended range of $\mathbf{k}$.

\vspace{0.05cm}
\begin{center}
{\bf C. Inputs from experiments on ultra-cold Fermi atoms}
\end{center}
\vspace{0.05cm}

The original motivation for looking at $A(\mathbf{k},\omega)$ has been the issue of ``preformed pairs'' in high-temperature (cuprate) superconductors, before the occurrence of the BCS-BEC crossover was explicitly demonstrated with ultra-cold Fermi atoms (cf., e.g., Ref.\cite{GPS-2008}).

In this context, the interest in the detailed shape of $A(k,\omega)$ above $T_{c}$ across the BCS-BEC crossover has considerably raised lately, after a new measurement technique was introduced to probe directly the single-particle excitations of a Fermi gas 
\cite{Jin-2008}.
Intensity maps were thus obtained for the single-particle excitation spectra, relating the single-particle energy to the wave vector.
More recently, new measurements performed over an extended temperature range above $T_{c}$ \cite{JILA-Cam-2010} have revealed a BCS-like dispersion with a characteristic ``back-bending'' close to $k_{F}$ which identifies a pseudo-gap energy scale and persists well above $T_{c}$.

This finding gives us motivations for extending the theoretical analysis of Ref.\cite{PPSC-2002} for the dispersions of the peaks of $A(k,\omega)$ across the unitary region, although the widths of the peaks can increase considerably with respect to the BCS side, reflecting the fact that quasi-particle excitations may be poorly defined.
Specifically, the combined experimental and theoretical analysis of Ref.\cite{Cam-JILA-2011} suggests us to concentrate our efforts in the coupling range approximately between $(k_{F} a_{F})^{-1} = 0$ and $(k_{F} a_{F})^{-1} = 0.4$. 

\vspace{0.05cm}
\begin{center}
{\bf D. Emergence of the contact in $A(k,\omega)$}
\end{center}
\vspace{0.05cm}

Yet, it was pointed out \cite{Randeria-2010} that the persistence of the back-bending for large $k$ ($\gg k_{F}$) is dominated by interaction effects that do not reflect the pseudo-gap close to $k_{F}$.
Rather, it is connected with the universal $k^{-4}$ tail of the wave-vector distribution $n(k)$ of a dilute Fermi gas, whose coefficient is given by the Tan's contact $C$ \cite{Tan-2008-I,Tan-2008-II}.

This property can be readily verified within the t-matrix approximation that we use to obtain $A(k,\omega)$. 
When $k^{2}/(2m)$ or $|\omega_{n}|$ are much larger than the energy scales $k_{B} T$ and $|\mu|$, in fact, the self-energy 
(\ref{Matsubara-self-energy}) can be approximated by:
\begin{equation}
\Sigma(\mathbf{k},\omega_{n}) \simeq - \frac{1}{2} \, n_{f}  \, \Gamma_{0}(\mathbf{k},\omega_{n})  \, - \, \Delta_{\infty}^{2} \, G_{0}(\mathbf{k},-\omega_{n}) \,\, .                                                                                                                           \label{approximate-Matsubara-self-energy}
\end{equation}

\noindent
Here,
\begin{equation}
n_{f} = 2 \, \int \! \frac{d\mathbf{k}}{(2 \pi)^{3}} \, n_{f}(\mathbf{k})            \label{density-zeroth-level}
\end{equation}

\noindent
with
\begin{equation}
n_{f}(\mathbf{k})  = k_{B} T  \, \sum_{n} \, e^{i \omega_{n} \eta} \, G_{0}(\mathbf{k},\omega_{n})     \label{n-k}
\end{equation}

\noindent
is the free density associated with $G_{0}$ for given $\mu$, and 
\begin{equation}
\Delta_{\infty}^{2} = \int \! \frac{d\mathbf{q}}{(2 \pi)^{3}} \, k_{B} T  \, \sum_{\nu} \, 
e^{i \Omega_{\nu} \eta} \, \Gamma_{0}(\mathbf{q},\Omega_{\nu})                                                                           \label{delta-infinity}
\end{equation}

\noindent
is the square of the high-energy scale introduced in Ref.\cite{PPS-2009} that was mentioned in the Introduction.
The two terms on the right-hand side of the approximate expression (\ref{approximate-Matsubara-self-energy}) originate from the singularities in the complex frequency plane of the single-particle Green's function $G_{0}$ and of the particle-particle ladder $\Gamma_{0}$, in the order, once the sum over the Matsubara frequency in the expression (\ref{Matsubara-self-energy}) of the fermionic self-energy is transformed into a contour integral.      

Analytic continuation $i \omega_{n} \rightarrow \omega + i \eta$ to the real frequency axis then results into the following approximate expression for large $k$:
\begin{equation}
A(k,\omega) \simeq \left( 1 - \frac{\Delta_{\infty}^{2}}{4 \, \xi_{\mathbf{k}}^{2}} \right) \, \delta(\omega - \xi_{\mathbf{k}}) 
+ \frac{\Delta_{\infty}^{2}}{4 \, \xi_{\mathbf{k}}^{2}} \,\, \delta(\omega + \xi_{\mathbf{k}})            \label{approximate-spectral-density}
\end{equation}

\noindent
which presents indeed a well-defined structure at the negative frequency $\omega = - \xi_{\mathbf{k}}$.
One obtains correspondingly:

\begin{equation}
n(k) = \int_{-\infty}^{+\infty} \! d \omega \, f(\omega) \, A(k,\omega) \simeq \frac{\Delta_{\infty}^{2}}{4 \, \xi_{\mathbf{k}}^{2}} \approx \frac{(m \, \Delta_{\infty})^{2}}{k^{4}}                                                                                      \label{approximate-n(k)}
\end{equation}

\noindent
yielding the relation $C = (m \, \Delta_{\infty})^{2}$ between the contact $C$ and $\Delta_{\infty}$, which will be extensively used below.

In practice, the structure of $A(k,\omega)$ for negative real $\omega$ at large $k$ is not delta-like but spreads over a sizable frequency range.
This difference from the approximate result (\ref{approximate-spectral-density}) stems from the non-commutativity between taking the analytic continuation and performing the large-$k$ expansion of the self-energy, as recalled in the Appendix.
Nevertheless, the actual structure of $A(k,\omega)$ for negative $\omega$ preserves the same \emph{total area} 
$\Delta_{\infty}^{2}/(4 \, \xi_{\mathbf{k}}^{2})$ found above in the expression (\ref{approximate-spectral-density}).

\vspace{0.05cm}
\begin{center}
{\bf E. Connecting the two energies $\Delta_{\mathrm{pg}}$ and $\Delta_{\infty}$}
\end{center}
\vspace{0.05cm}

With these premises, it seems natural to frame the low-energy scale $\Delta_{\mathrm{pg}}$ and the high-energy scale $\Delta_{\infty}$ into a 
unified physical picture, in which they emerge from $A(k,\omega)$ in the two distinct ranges of wave vectors $k \approx k_{F}$ and 
$k \gg k_{F}$, respectively.
To this end, we shall extend to the unitary limit and beyond the analysis that was limited in Ref.\cite{PPSC-2002} to the BCS side of the unitary region, by following the dispersions, weights, and widths of the peaks of $A(k,\omega)$ from $k = 0$, through $k \approx k_{F}$, and up to $k \gg k_{F}$, even in cases when these peaks appear quite broad and overlapping.

On physical grounds, the evolution, from $\Delta_{\mathrm{pg}}$ when $k \approx k_{F}$ to $\Delta_{\infty}$ when $k \gg k_{F}$, is expected on the basis of a (local in space and transient in time) order which is established by pair fluctuations above $T_{c}$, in the absence of long-range order (as described by mean field below $T_{c}$).
This is in line with the definition of the contact $C$ (and thus of $\Delta_{\infty}$) through the short-range behavior of the pair-correlation function between opposite spins \cite{Tan-2008-I}, which corresponds to $k \gg k_{F}$; while the pseudo-gap $\Delta_{\mathrm{pg}}$ is expected to depend on pair-correlations that are established more extensively over medium range, which corresponds to  $k \approx k_{F}$.
 
As a consequence, we expect the ``pseudo-gap physics'' to be associated with pair correlations which are built at intermediate distances of the order of $k_{F}^{-1}$, while the ``contact physics'' with pair correlations that survive even at smaller distances ($\ll k_{F}^{-1}$).
Both quantities $\Delta_{\mathrm{pg}}$ and $\Delta_{\infty}$ are thus affected by the same sort of pair correlations in the particle-particle channel, to which the t-matrix approximation that we adopt in this paper provides an important contribution. 
This is because the particle-particle ladder propagator [given by Eq.(\ref{particle-particle-ladder}) in the attractive case and by Eq.(\ref{separable-particle-particle-ladder}) in the repulsive case] generalizes to a many-body environment the two-body t-matrix, which describes two-body binding 
and scattering of unbound particles at the same time. 

Within this approximation, (the square of) $\Delta_{\infty}$ is defined from Eq.(\ref{delta-infinity}) as a wave-vector and frequency averaging of the particle-particle ladder propagator.
The wave-vector and frequency structures of the same propagator give also rise to the pseudo-gap $\Delta_{\mathrm{pg}}$, which then emerges as a characteristic low-energy scale in the single-particle excitations.

Consistently with this physical picture of locally established pair correlations, we expect $\Delta_{\mathrm{pg}}$ to survive above
$T_{c}$ over a more limited temperature range than $\Delta_{\infty}$, since thermal fluctuations act first to destroy the order established over intermediate distances. 
Our analysis about the characteristic features of the spectral functions will accordingly be extended over a meaningful temperature interval above $T_{c}$.

\vspace{0.05cm}
\begin{center}
{\bf F. Working procedures}
\end{center}
\vspace{0.05cm}

Working experience on the single-particle spectral function suggests us to identify the low-energy scale $\Delta_{\mathrm{pg}}$ in the range $0 \lesssim k  \lesssim 2 k_{F}$, while the high-energy scale $\Delta_{\infty}$ can be extracted with sufficient accuracy already from the not too extreme range $2 k_{F} \lesssim k  \lesssim 4 k_{F}$.
In a more extreme range of $k$ ($\gtrsim 4 k_{F}$), in fact, it would become quite difficult to determine $A(k,\omega)$ for large negative $\omega$. 

Owing to the shape of the $\omega$-structures of $A(k,\omega)$ for given $k$, different strategies need to be adopted in the above
two ranges of wave vectors.
Namely, in the range $0 \lesssim k  \lesssim 2 k_{F}$ the shape of the two peaks of $A(k,\omega)$ permits, in practice, to both follow their dispersions for varying $k$ and identify their weights in terms of the total area they comprise.
In the range $2 k_{F} \lesssim k  \lesssim 4 k_{F}$, on the other hand, the structure of $A(k,\omega)$ at negative frequencies is so broad that only its total area can be reasonably identified.
\begin{center}
{\bf Range $\mathbf{0 \lesssim k  \lesssim 2 k_{F}}$:}
\end{center}

Let's first consider the range $0 \lesssim k  \lesssim 2 k_{F}$. 
Here, the \emph{dispersions} that we are able to determine independently for the two peaks at positive and negative frequencies are fitted, respectively, by the BCS-like expressions:
\begin{equation}
\omega_{(\pm)}(k) \, = \, \pm \, \sqrt{ \left( \frac{k^{2}}{2m} \, - \, \frac{k_{L (\pm)}^{2}}{2m} \right)^{2}
\, + \, \Delta_{\mathrm{pg} (\pm)}^{2} }                                                                                                \label{dispersion-relations}
\end{equation}

\noindent
where a different pseudo-gap energy $\Delta_{\mathrm{pg}}$ is introduced for the upper $(+)$ and lower $(-)$ branches.
The fitting identifies, in addition, the locations $k_{L (+)}$ for the ``up-bending" of the upper branch and $k_{L (-)}$ for the ``down-bending" of the lower branch \cite{footnote-1}.
The fittings are carried out by a $\chi^{2}$-analysis.
We have found that $k_{L (+)}$ is consistently smaller than $k_{L (-)}$ in all situations we have examined.

Because the peaks at negative and positive frequencies are in general broad and partially overlapping with each other, we have chosen to determine operatively their \emph{weights} at a given $k$ by integrating $A(k,\omega)$ from $- \infty$ up to $\omega = 0$ and from $\omega = 0$ up to $+ \infty$, in the order.
We then fit these values obtained for several $k$ by the BCS-like expressions:
\begin{equation}
v(k)^{2} \, = \, \frac{1}{2} \, \left( 1 \, - \, \frac{ \frac{k^{2}}{2m} \, - \, \mu_{\mathrm{eff}}}
{\sqrt{ \left( \frac{k^{2}}{2m} \, - \, \mu_{\mathrm{eff}} \right)^{2} \, + \, \Delta_{\mathrm{pg} (-)}^{2} } }  \right)   \label{BCS-weight-v2}
\end{equation}

\noindent
for the lower branch at negative frequencies, and
\begin{equation}
u(k)^{2} \, = \, \frac{1}{2} \, \left( 1 \, + \, \frac{ \frac{k^{2}}{2m} \, - \, \mu_{\mathrm{eff}}}
{\sqrt{ \left( \frac{k^{2}}{2m} \, - \, \mu_{\mathrm{eff}} \right)^{2} \, + \, \Delta_{\mathrm{pg} (+)}^{2} } }  \right)   \label{BCS-weight-u2}
\end{equation}

\noindent
for the upper branch at positive frequencies.
Here, $\Delta_{\mathrm{pg} (\pm)}$ are the pseudo-gap energies already determined from the fitting (\ref{dispersion-relations}) to the dispersions, and $\mu_{\mathrm{eff}}$ is a new parameter common to the two branches which is determined from the position where the weights cross.
We shall find that the value of $\sqrt{2 m  \mu_{\mathrm{eff}}}$ is intermediate between $k_{L (+)}$ and $k_{L (-)}$ obtained from the dispersions 
(\ref{dispersion-relations}) of the two branches, consistently with enforcing a common value $\mu_{\mathrm{eff}}$ in 
Eqs.(\ref{BCS-weight-v2}) and (\ref{BCS-weight-u2}).

Note that the expressions (\ref{BCS-weight-v2}) and (\ref{BCS-weight-u2}) do not require \emph{a priori} the sum 
$u(k)^{2} + v(k)^{2}$ to be unity, as it would be the case for the coherence factors in a strict BCS description.
Deviations of the sum of the expressions (\ref{BCS-weight-v2}) and (\ref{BCS-weight-u2}) from unity can thus be taken as a test for the validity of the effective BCS description that we are attempting to establish in the normal phase close to $T_{c}$.

Concerning, finally, the \emph{widths} of the peaks of $A(k,\omega)$, they will be conventionally determined as the full widths at half maximum.
Whenever necessary, however, these values will be compared with the numerical values obtained alternatively by a two-Lorentzian fit to the peaks of $A(k,\omega)$, a procedure which is of course more reliable the more these peaks are separated in frequency.

The BCS-like expressions (\ref{dispersion-relations}), (\ref{BCS-weight-v2}), and (\ref{BCS-weight-u2}) are meant to test the persistence of a BCS-like description in the normal phase of a Fermi gas with an attractive pairing interaction.
In addition, the presence in Eq.(\ref{dispersion-relations}) of the ``Luttinger wave vectors'' $k_{L (\pm)}$ highlights the persistence of an underlying Fermi surface for the single-particle excitations, which represents the last remnant of what would be a Fermi-liquid description of the Fermi gas if attractive pairing interactions above $T_{c}$ would not be considered \cite{Cam-JILA-2011}.
Nevertheless, the large widths associated with the peaks of $A(k,\omega)$ represent \emph{per se} an evidence that a Fermi-liquid 
description above $T_{c}$ does not apply in the presence of pairing fluctuations.
\begin{center}
{\bf Range $\mathbf{ 2 k_{F} \lesssim k  \lesssim 4 k_{F}}$:}
\end{center}

In this range, the structure at negative frequencies in $A(k,\omega)$ becomes so spread and broad that it is meaningless to determine its dispersion numerically and then try to fit it by an expression similar to (\ref{dispersion-relations}).
In this case, however, it remains meaningful to determine the total area of the broad structure at negative $\omega$ over a chosen mesh of $k$ and 
then make a $\chi^{2}$-fit to these values through the following expression which is inspired by 
Eq.(\ref{approximate-spectral-density}): 
\begin{equation}
v(k)_{\mathrm{large}}^{2} \, = \, \frac{\Delta_{\mathrm{large}}^{2}}{4 \, \left( \frac{k^{2}}{2m} \, - \, \mu_{\mathrm{large}} \right)^{2}}         
                                                                                 \label{v-2-large}
\end{equation}

\noindent
where $\Delta_{\mathrm{large}}$ and $\mu_{\mathrm{large}}$ are fitting parameters to be determined in this range of ``large'' $k$.

In practice, it is convenient to set $\mu_{\mathrm{large}}$ at the corresponding value of the thermodynamic chemical potential $\mu$ from the outset (thus leaving $\Delta_{\mathrm{large}}$ as the only fitting parameter).
This is because in the coupling range of interest $\mu_{\mathrm{large}}$ is small enough that it becomes meaningless to extract it from the denominator of Eq.(\ref{v-2-large}) where $k^{2}/(2m)$ dominates in this range of $k$.

We will check whether the value of $\Delta_{\mathrm{large}}$ determined in this way coincides with the value of $\Delta_{\infty}$ obtained independently by the expression (\ref{delta-infinity}), and how it differs from the value $\Delta_{\mathrm{pg} (-)}$ of the pseudo-gap 
obtained above near the region of the back-bending of the lower branch.

\vspace{0.05cm}
\begin{center}
{\bf G. Results for dispersions, weights, and widths}
\end{center}
\vspace{0.05cm}

\begin{figure}[h]
\begin{center}
\includegraphics[angle=0,width=5.5cm]{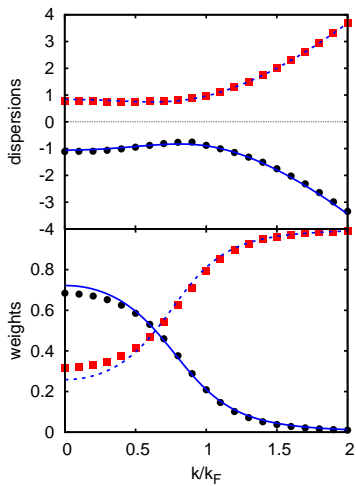}
\caption{Dispersions (upper panel) and corresponding weights (lower panel) at unitarity and $T = T_{c}$.
Circles (squares) and full (dashed) lines represent the results of the numerical calculation and 
of the BCS-like fits for the lower (upper) branch at negative and positive frequencies, respectively. Energies are in units of $E_{F}$.}
\label{fig-II-3}
\end{center}
\end{figure}

We pass to determine the quantities of interest according to the procedures outlined above.
We shall specifically consider the two coupling values $(k_{F} a_{F})^{-1} = 0$ and $(k_{F} a_{F})^{-1} = 0.25$ as representatives
of the coupling range where pseudo-gap phenomena are expected to be maximal.
Two representative temperatures will also be considered for each coupling.

Figure~\ref{fig-II-3} shows the dispersion relations and weights of the two peaks of $A(k,\omega)$ at $T_{c}$ when
$(k_{F} a_{F})^{-1} = 0$ in the range $0 \le k  \le 2 k_{F}$, as obtained from the numerical calculation based on 
Eq.(\ref{single-particle-spectral-function}) and from the fits obtained according to Eqs.(\ref{dispersion-relations})-(\ref{BCS-weight-u2}).
In this case the fitting parameters are
$k_{L (-)} = 0.78 k_{F}$ and $\Delta_{\mathrm{pg} (-)} = 0.83 E_{F}$ for the lower branch, 
and $k_{L (+)} = 0.62 k_{F}$ and $\Delta_{\mathrm{pg} (+)} = 0.74 E_{F}$ for the upper branch, 
while $\mu_{\mathrm{eff}} = 0.41E_{F}$ is quite close to the corresponding value of the thermodynamic potential 
$\mu = 0.365 E_{F}$ obtained within the t-matrix approximation. 
[The results of the numerical calculations have already been reported in the central panels of Fig.~\ref{fig-II-1}, although with the different purpose of comparing them with the mean-field description at $T=0$.]

In this case the BCS-like fits are excellent for the dispersions and quite good for the weights, especially near the value
$k = 0.64 k_{F}$ where the weights exchange with one another.
Note also that, while the numerical values of the weights for each value of $k$ are specular to each other about one half, the fitted values are not always so indicating deviations from their sum being unity.

\begin{figure}[h]
\begin{center}
\includegraphics[angle=0,width=5.5cm]{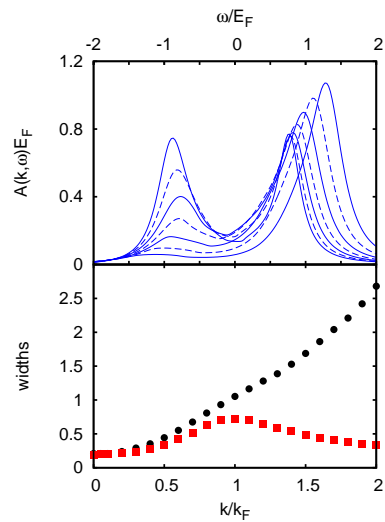}
\caption{Unitarity limit and $T = T_{c}$. 
Upper panel: $A(k,\omega)$ vs $\omega$ for the wave vectors $k = (0.6,0.7,0.8,0.9,1.0,1.1,1.2) k_{F}$ corresponding to the peaks at negative frequency from top to bottom (here full and dashed lines alternate to help the analysis of the figure).
Lower panel: Widths (in units of $E_{F}$) of the peaks at negative (circles) and positive (squares) frequencies.}
\label{fig-II-4}
\end{center}
\end{figure}

This success of a BCS-like interpretation for the dispersions and weights of the peaks should be complemented by the further information about their widths.
This is done in Fig.~\ref{fig-II-4}, where in the upper panel the shape of $A(k,\omega)$ at $T_{c}$ and $(k_{F} a_{F})^{-1} = 0$ is shown explicitly for several wave vectors, while in the lower panel the corresponding widths of the peaks at negative and positive frequencies are reported over a wider set of wave vectors. 

In all cases, the widths are rather large (being comparable to $E_{F}$) and show strong deviations from what would be expected for a Fermi liquid picture, according to which they should acquire a minimum value at about $k_{F}$.
These deviations from a Fermi liquid picture are of course expected for a Fermi gas with attractive interaction, taking further into account 
that at unitarity the value of $T_{c}$ is a considerable fraction of the Fermi temperature $T_{F}$ whereas a Fermi-liquid description holds only for $T \ll T_{F}$ \cite{Fermi-liquids-books}.

\begin{figure}[h]
\begin{center}
\includegraphics[angle=0,width=5.5cm]{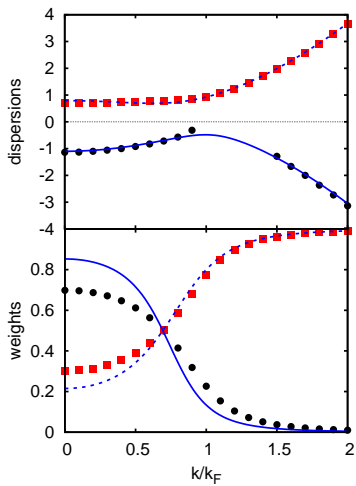}
\caption{Dispersions (upper panel) and weights (lower panel) at unitarity and $T = 1.2 \, T_{c}$.
Conventions are as in Fig.~\ref{fig-II-3}.}
\label{fig-II-5}
\end{center}
\end{figure}

The above analysis of pseudo-gap phenomena around $k_{F}$ is expected to remain meaningful for temperatures larger than $T_{c}$, but not exceeding the pair-breaking temperature scale $T^{*}$ where a ``preformed-pair scenario'' is bound to fade away.
In particular, at unitarity the value of $T^{*}$ (as estimated by the mean-field critical temperature) is about twice the value of $T_{c}$ given by the t-matrix approximation we are considering \cite{Randeria-1993}\cite{PPPS-2004}.
As a representative case of a temperature above $T_{c}$, Fig.~\ref{fig-II-5} shows the dispersions and weights obtained from $A(k,\omega)$ at unitarity and $T = 1.2 \, T_{c}$.
The fitting parameters are now
$k_{L (-)} = 0.99 k_{F}$ and $\Delta_{\mathrm{pg} (-)} = 0.48 E_{F}$ for the lower branch, 
and $k_{L (+)} = 0.69 k_{F}$ and $\Delta_{\mathrm{pg} (+)} = 0.62 E_{F}$ for the upper branch, 
while $\mu_{\mathrm{eff}} = 0.48E_{F}$ (to be compared with the thermodynamic value $\mu = 0.39E_{F}$).

Note that in this case the analysis of the dispersion of the lower branch had to be interrupted over a non-negligible interval of $k$ about $k_{F}$, because in this interval the structure of $A(k,\omega)$ at negative frequencies is almost completely masked by the stronger structure at positive frequencies.
This represents a signal that pseudo-gap phenomena are beginning to fade away at this temperature.
We have nevertheless performed a BCS-like fit to the part of the dispersion that can still be clearly identified, as shown by the full curve in the upper panel of Fig.~\ref{fig-II-5}.
By our procedure no problem instead arises in identifying the corresponding weights reported in the lower panel of Fig.~\ref{fig-II-5}, which follow again a BCS-like dispersion although with less accuracy than those shown at $T_{c}$ in Fig.~\ref{fig-II-3}.

\begin{figure}[h]
\begin{center}
\includegraphics[angle=0,width=5.5cm]{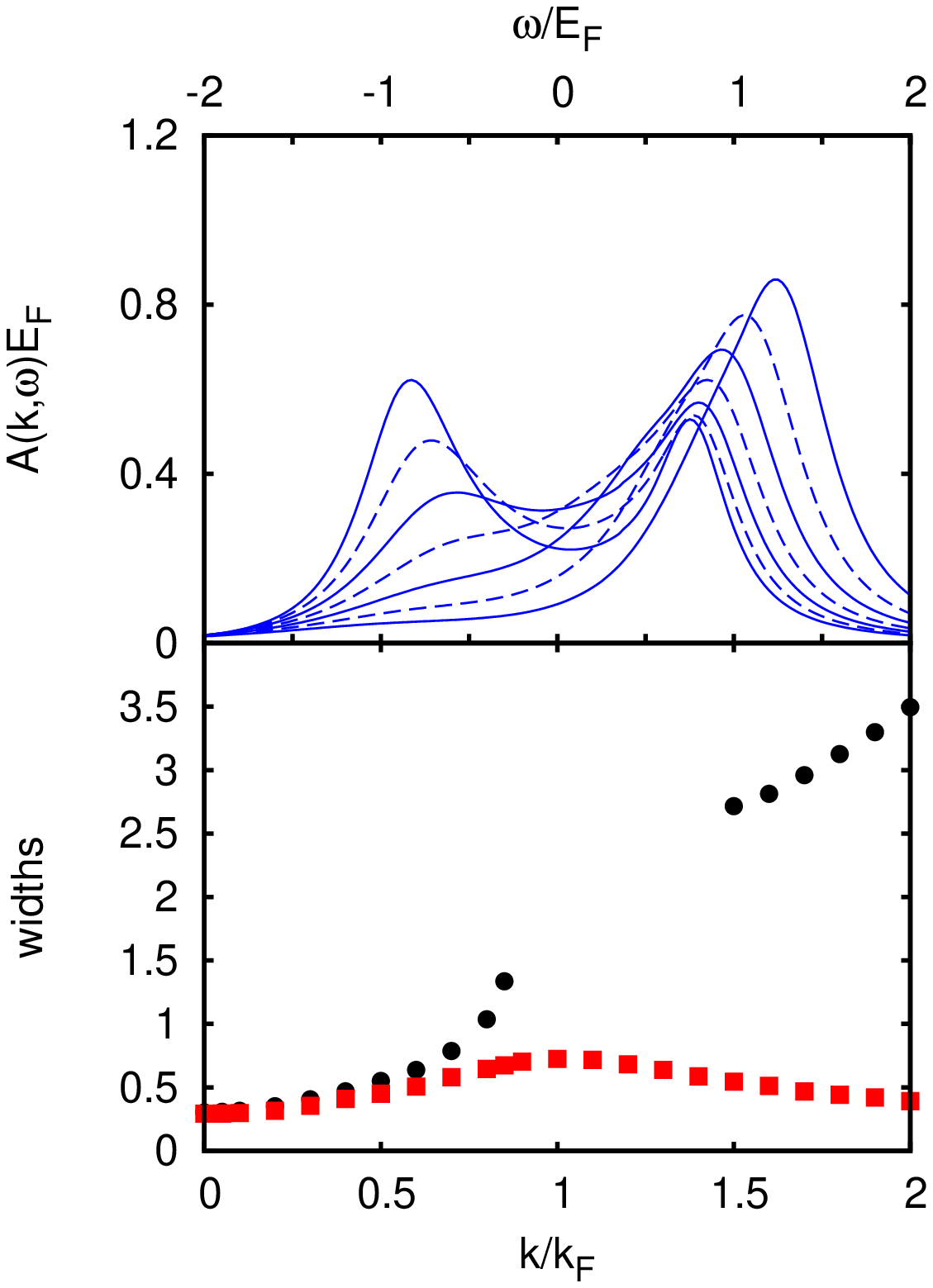}
\caption{Unitarity limit and $T = 1.2 \, T_{c}$. 
Upper panel: $A(k,\omega)$ vs $\omega$ for the same wave vectors as in the upper panel of Fig.~\ref{fig-II-4}.
Lower panel: Widths (in units of $E_{F}$) of the peaks at negative (circles) and positive (squares) frequencies.}
\label{fig-II-6}
\end{center}
\end{figure}

The corresponding shapes of $A(k,\omega)$ vs $\omega$ for a chosen set of $k$ across $k_{F}$ are shown explicitly in the upper panel of Fig.~\ref{fig-II-6}, from which one can appreciate the phenomenon mentioned above, when the structure of $A(k,\omega)$ for the lower branch becomes a shoulder attached to the structure of the upper branch.
The corresponding broadenings of these two structures are reported in the lower panel of Fig.~\ref{fig-II-6}, which reinforces our conclusion about the non-Fermi-liquid nature of the system.

A question naturally arises, about whether or not these profiles of $A(k,\omega)$ still allow one to identify the presence of a pseudo-gap in the crucial range of wave vectors about $k_{F}$.
As the upper panel of Fig.~\ref{fig-II-5} shows, in fact, an \emph{overall} BCS-like fit to the lower branch can be attempted even in this case, because the two structures of $A(k,\omega)$ remain distinct from each other away from $k_{F}$.

\begin{figure}[t]
\begin{center}
\includegraphics[angle=0,width=6.5cm]{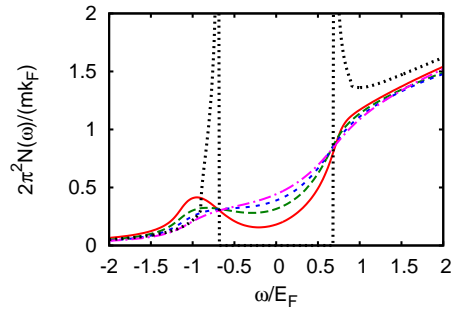}
\caption{Density of states per spin component vs $\omega$ calculated at unitarity within the t-matrix approximation for the temperatures: 
$T = T_{c}$ (full line), $T = 1.2 \, T_{c}$ (long-dashed line), $T = 1.4 \, T_{c}$ (short-dashed line), and 
$T = 1.65 \, T_{c}$ (dot-dashed line). In the present case, $T^{*} \approx 2 \, T_{c}$.
The dotted line shows the corresponding mean-field result when $T = 0$.
The non-interacting value $m k_{F}/(2 \pi^{2})$ of $N(\omega = 0)$ at $T = 0$ is used for normalization.}
\label{fig-II-7}
\end{center}
\end{figure}

The relevance of this restricted interval about $k_{F}$ can be strongly reduced by averaging the profiles of $A(k,\omega)$ over \emph{all} wave vectors, in the way it is done in the definition (\ref{DOS}) of the single-particle density of states $N(\omega)$.
Figure ~\ref{fig-II-7} shows a plot of $N(\omega)$ vs $\omega$ at unitarity for several temperatures at and above $T_{c}$ \cite{footnote-3}.
For increasing $T$, the depression of $N(\omega)$ near $\omega = 0$ well survives for $T = 1.2 \, T_{c}$ at which
the dispersion of the lower branch near $k_{F}$ in Fig.~\ref{fig-II-5} had to be interrupted, and progressively disappears for temperatures somewhat below the pair-breaking temperature scale $T^{*}$.
That the depression of density of states survives at temperatures higher than the crossover temperature where the pseudo-gap features disappear in the spectral function was previously discussed in Refs.\cite{Levin-1997,Ohashi-2009}.   
At about $T^{*}$, $N(\omega)$ for $\omega = 0$ coincides (within a few percent) with its non-interacting value evaluated at the same temperature and chemical potential, indicating that all effects of pairing have faded away at $\omega = 0$ (although they will persist at higher temperatures for $\omega \ll - E_{F}$, indicating the survival of the ``contact'' even at quite high temperatures \cite{PPPS-2010}).  
The density of states obtained within mean field at zero temperature is also reported for comparison in Fig.~\ref{fig-II-7}, and shows two sharp peaks located at $\pm \Delta$ with $\Delta = 0.69 \, E_{F}$.

\begin{figure}[t]
\begin{center}
\includegraphics[angle=0,width=5.5cm]{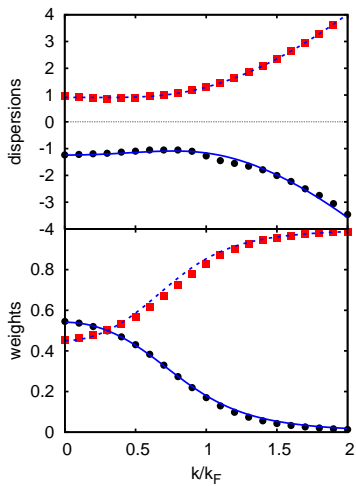}
\caption{Dispersions (upper panel) and weights (lower panel) at $T = T_{c}$ for the coupling $(k_{F} a_{F})^{-1} = 0.25$.
Conventions are as in Fig.~\ref{fig-II-3}.}
\label{fig-II-8}
\end{center}
\end{figure}

It is important to extend the above analysis past the unitarity limit to the BEC side of the crossover (but still before the pseudo-gap turns into a real gap associated with the binding energy of the composite bosons which form in the BEC limit).
To this end, Fig.~\ref{fig-II-8} shows the dispersions and weights at $T_{c}$ for the coupling $(k_{F} a_{F})^{-1} = 0.25$, together with the corresponding BCS-like fits where now $k_{L (-)} = 0.77 k_{F}$ and $\Delta_{\mathrm{pg} (-)} = 1.09 E_{F}$ for the lower branch, and
$k_{L (+)} = 0.28 k_{F}$ and $\Delta_{\mathrm{pg} (+)} = 0.91 E_{F}$ for the upper branch, 
while $\mu_{\mathrm{eff}} = 0.09E_{F}$ (which in this case almost coincides with the thermodynamic value).
Compared with Fig.~\ref{fig-II-3}, the dispersions have now become quite flat in the range $ 0 < k  \lesssim k_{F}$, while the two weights cross at smaller value of $k$.
At even stronger couplings, the lower dispersion down-bends and the upper dispersion up-bends already at $k = 0$ while the weights always remain well separated from each other for all $k$ (as it is shown in the lower panels of Fig.~\ref{fig-II-1} for the coupling 
$(k_{F} a_{F})^{-1} = 1.0$).

\begin{figure}[h]
\begin{center}
\includegraphics[angle=0,width=5.5cm]{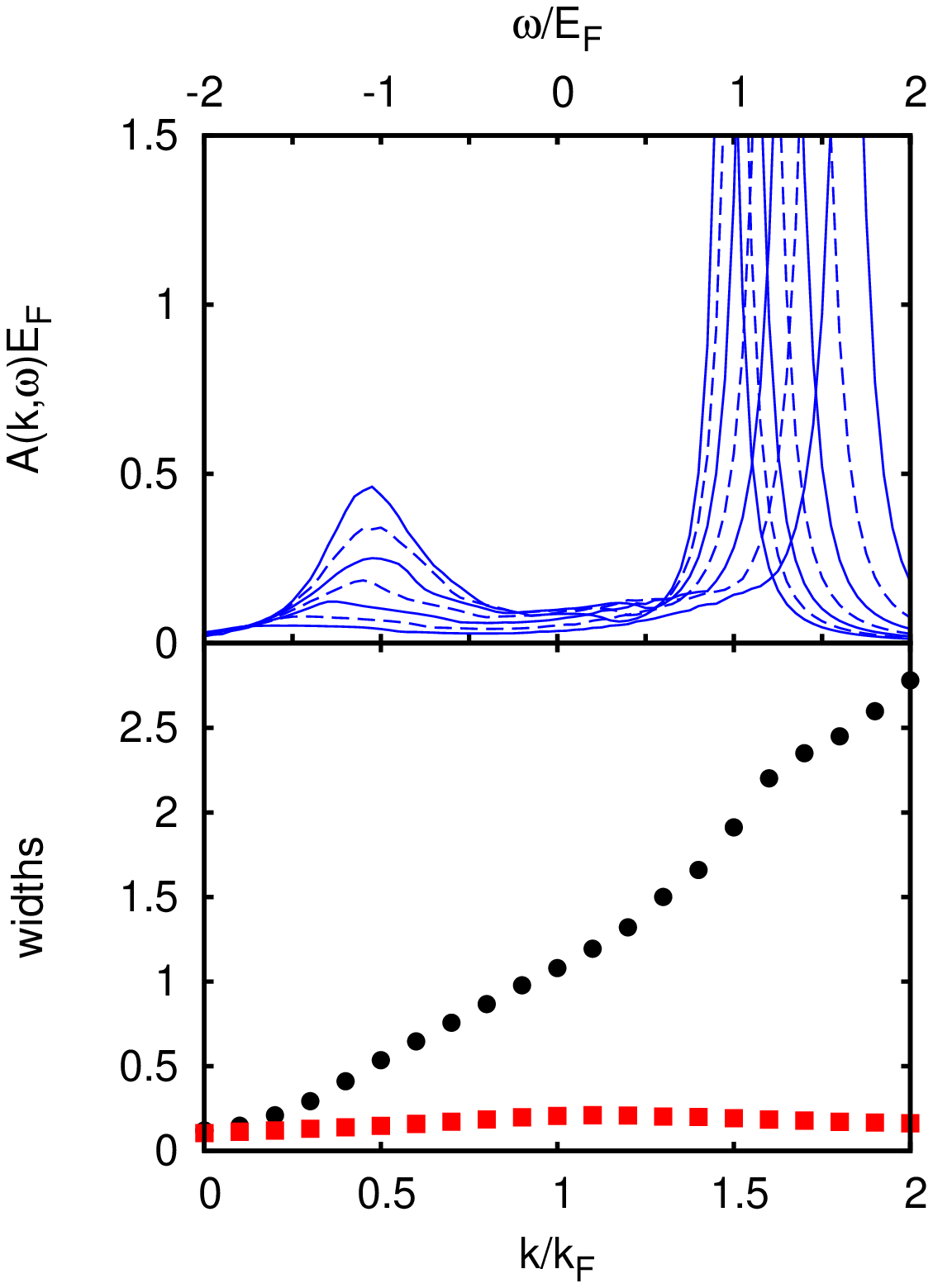}
\caption{Coupling $(k_{F} a_{F})^{-1} = 0.25$ and $T = T_{c}$. 
Upper panel: $A(k,\omega)$ vs $\omega$ for the same wave vectors as in the upper panel of Fig.~\ref{fig-II-4}.
Lower panel: Widths (in units of $E_{F}$) of the peaks at negative (circles) and positive (squares) frequencies.}
\label{fig-II-9}
\end{center}
\end{figure} 

For completeness, Fig.~\ref{fig-II-9} shows the corresponding shapes of $A(k,\omega)$ across $k_{F}$ (upper panel)
as well as the broadenings of two structures of $A(k,\omega)$ (lower panel).

\begin{figure}[t]
\begin{center}
\includegraphics[angle=0,width=5.5cm]{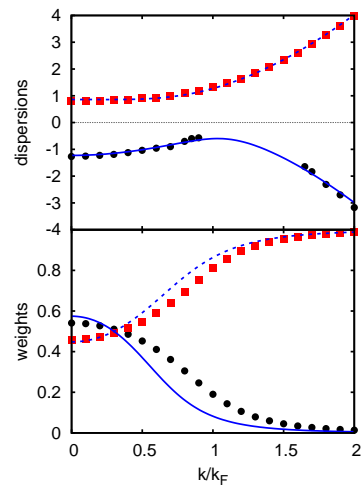}
\caption{Dispersions (upper panel) and weights (lower panel) at $T = 1.4 \, T_{c}$ for the coupling $(k_{F} a_{F})^{-1} = 0.25$.
Conventions are as in Fig.~\ref{fig-II-3}.}
\label{fig-II-10}
\end{center}
\end{figure}

\begin{figure}[h]
\begin{center}
\includegraphics[angle=0,width=5.5cm]{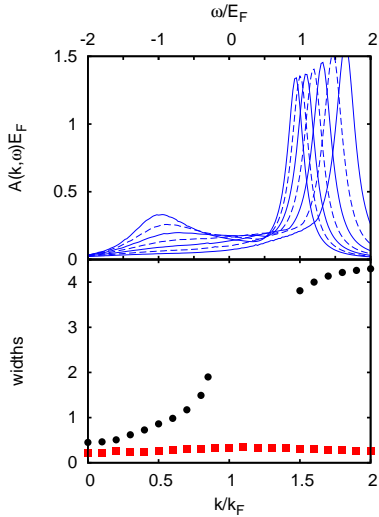}
\caption{Coupling $(k_{F} a_{F})^{-1} = 0.25$ and $T = 1.4 \, T_{c}$. 
Upper panel: $A(k,\omega)$ vs $\omega$ for the same wave vectors as in the upper panel of Fig.~\ref{fig-II-4}.
Lower panel: Widths (in units of $E_{F}$) of the peaks at negative (circles) and positive (squares) frequencies.}
\label{fig-II-11}
\end{center}
\end{figure}

When the coupling increases toward the BEC regime, the pair-breaking temperature $T^{*}$ increases more markedly than $T_{c}$ 
\cite{Randeria-1993}\cite{PPPS-2004} and pairing fluctuations are accordingly expected to affect $A(k,\omega)$ over a progressively wider temperature range 
above $T_{c}$.
We then report in Fig.~\ref{fig-II-10} the dispersions and widths of the two branches of $A(k,\omega)$ for the coupling 
$(k_{F} a_{F})^{-1} = 0.25$ and the higher temperature $T = 1.4 \, T_{c}$.
Again, a signal that the pseudo-gap is beginning to fade away emerges from the analysis of the dispersion for the lower branch, which has to be interrupted about $k_{F}$.
The fitting parameters are now
$k_{L (-)} = 1.03 k_{F}$ and $\Delta_{\mathrm{pg} (-)} = 0.60 E_{F}$ for the lower branch, 
and $k_{L (+)} = 0.25 k_{F}$ and $\Delta_{\mathrm{pg} (+)} = 0.86 E_{F}$ for the upper branch, 
while $\mu_{\mathrm{eff}} = 0.09E_{F}$ (to be compared with the thermodynamic value $\mu = 0.14 E_{F}$).
The corresponding shapes of $A(k,\omega)$ across $k_{F}$ and the broadenings of two structures of $A(k,\omega)$ are shown, respectively, in the upper and lower panels of Fig.~\ref{fig-II-11}.

\begin{figure}[t]
\begin{center}
\includegraphics[angle=0,width=5.2cm]{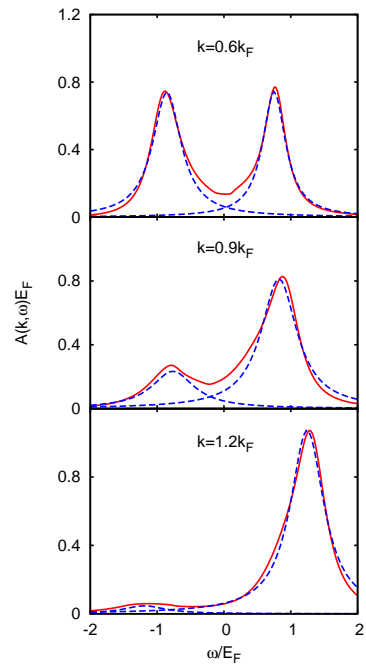}
\caption{Two-Lorentzian fits (dashed lines) of $A(k,\omega)$ vs $\omega$ (full lines) at unitarity and $T = T_{c}$, when 
$k/k_{F} = (0.6,0.9,1.2)$ from top to bottom.}
\label{fig-II-12}
\end{center}
\end{figure}

Beginning with Fig.~\ref{fig-II-2}, we have often emphasized that one of the major characteristics of the two structures of $A(k,\omega)$ (at and) above $T_{c}$ is their substantial broadening, which may hinder in practice a straightforward identification of the pseudo-gap about $k_{F}$ in cases when these structures strongly overlap with each other.
In these cases, however, one may resort to a two-Lorentzian fit of the two structures of $A(k,\omega)$ which helps separating them.
This is shown Fig.~\ref{fig-II-12} where the dashed lines represent the two Lorentzians.
For instance, by this type of fit our previous estimates for the weight ($0.29$) and width ($0.93 E_{F}$) of the structure of $A(k,\omega)$ 
at $k = 0.9 k_{F}$ corresponding to the lower branch are replaced by $0.28$ and $0.77 E_{F}$, in the order.
Comparable deviations are obtained in the other cases.
These results thus confirm the validity of our previous analysis where the weights and widths were extracted from $A(k,\omega)$ in a simpler fashion.

\begin{figure}[t]
\begin{center}
\includegraphics[angle=0,width=8.7cm]{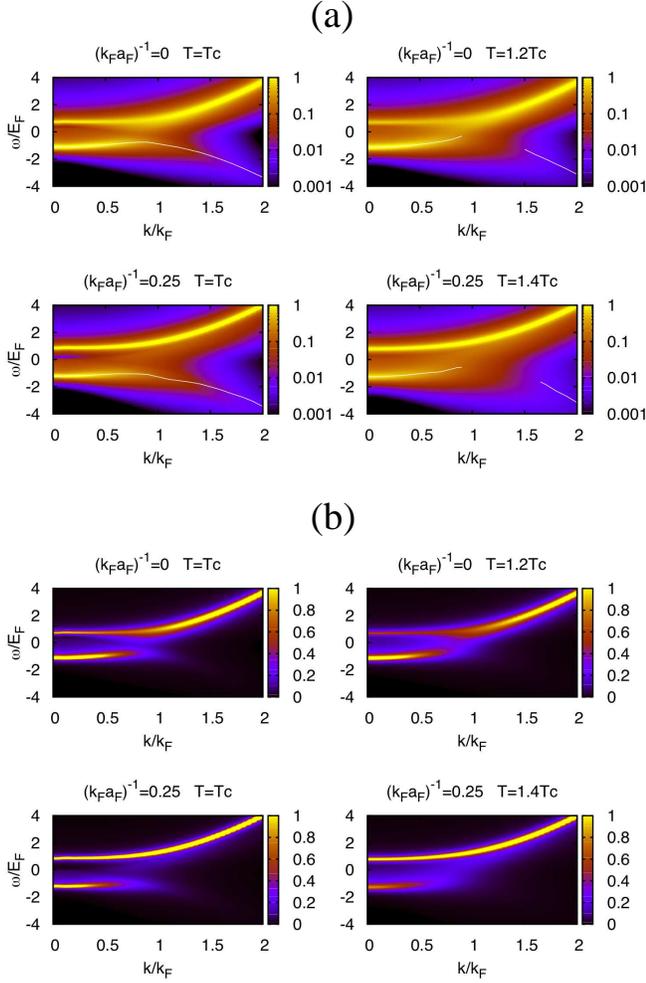}
\caption{Intensity plots for the single-particle spectral function at given temperature and coupling, in (a) logarithmic and (b) linear scale.
The thin white line identifies the dispersion of the lower branch as determined in the previous figures.}
\label{fig-II-12-bis}
\end{center}
\end{figure}

The numerical values of the dispersions, weights, and widths that were reported in the previous figures were all obtained from the detailed profiles of the single-particle spectral function $A(k,\omega)$, which were also shown in the same figures.
It may also be of use, however, to organize the spectra of $A(k,\omega)$ for a range of $k$ and $\omega$ into a single intensity plot.
This is done in Fig.~\ref{fig-II-12-bis}(a) for the same set of temperatures and couplings considered in the previous figures.
Similar intensity plots were presented in Refs.\cite{Ohashi-2009} and \cite{Levin-2010}. 
Note that the log scale, used here like in the experimental works \cite{Jin-2008,JILA-Cam-2010}, makes the back-bending more evident when compared 
with the intensity plots presented in Refs.\cite{Ohashi-2009} and \cite{Levin-2010}.
For the sake of comparison with those references, we also report in Fig.~\ref{fig-II-12-bis}(b) the same intensity plots in a linear scale.

Thus far we have concentrated our attention to the range $0 \lesssim k \lesssim 2 k_{F}$ where the pseudo-gap physics manifests itself.
We pass now to discuss the more asymptotic range $2 k_{F} \lesssim k \lesssim 4 k_{F}$ where the contact physics emerges.
To this end, we adopt the procedure outlined in sub-section II-F and determine the parameter $\Delta_{\mathrm{large}}$ from the expression
(\ref{v-2-large}) with $\mu_{\mathrm{large}}$ fixed at the corresponding value of the chemical potential.

\begin{figure}[t]
\begin{center}
\includegraphics[angle=0,width=8.5cm]{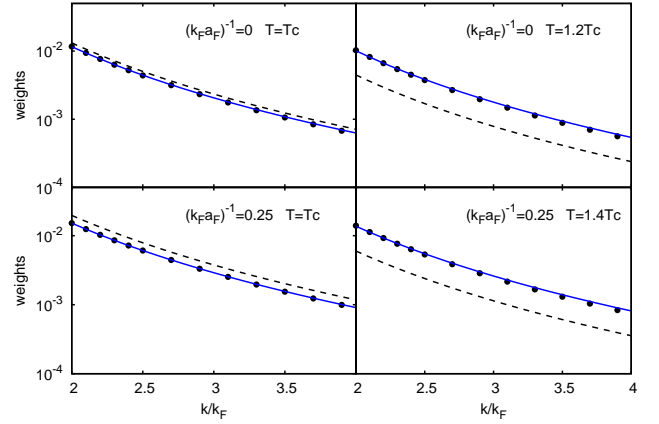}
\caption{Weights of $A(k,\omega)$ at negative $\omega$ (circles) are fitted according to the expression discussed in sub-section II-F (full lines)
for different couplings and temperatures. Dashed lines correspond to what would be obtained by using in that expression the numerical values of the low-energy scale $\Delta_{\mathrm{pg}(-)}$.}
\label{fig-II-13}
\end{center}
\end{figure}

Figure~\ref{fig-II-13} shows the weights of the structure of $A(k,\omega)$ at negative $\omega$ for the values of coupling and temperature considered so far, as determined numerically (circles) over a mesh of values of $k$ in the range $2 k_{F} \le k \le 4 k_{F}$ and then fitted
(full lines) in terms of the expression (\ref{v-2-large}).

These fits are also compared with an expression of the form (\ref{v-2-large}), where now the low-energy scale $\Delta_{\mathrm{pg}(-)}$ that was previously determined in the range $0 \lesssim k \lesssim 2 k_{F}$ replaces $\Delta_{\mathrm{large}}$ (dashed lines).
The appreciable deviations from the numerical values of the weights that result show that the high-energy scale 
$\Delta_{\infty}$ can be distinguished from the low-energy scale $\Delta_{\mathrm{pg}(-)}$ by inspecting the shape of $A(k,\omega)$ in different ranges of $k$.
Note in particular that, as soon the temperature is increased above $T_{c}$, $\Delta_{\mathrm{pg}(-)}$ becomes rapidly smaller than 
$\Delta_{\infty}$.
This is consistent with our expectation that $\Delta_{\infty}$, being associated with local pair correlations of shorter range with respect to $\Delta_{\mathrm{pg}}$, survives at higher temperatures.

\begin{figure}[t]
\begin{center}
\includegraphics[angle=0,width=6.0cm]{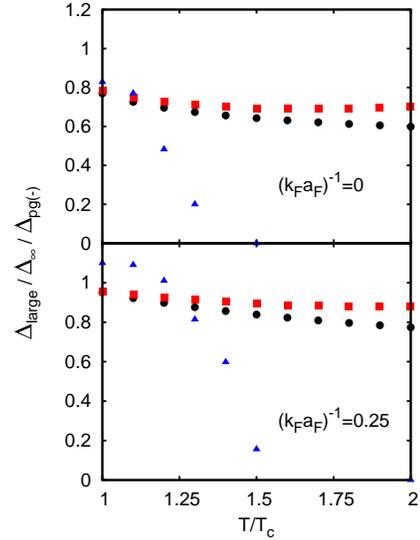}
\caption{Temperature dependence of $\Delta_{\mathrm{large}}$ (squares), $\Delta_{\infty}$ (circles), and $\Delta_{\mathrm{pg}(-)}$ (triangles)
              in units of $E_{F}$ for the couplings $(k_{F} a_{F})^{-1} = 0$ (upper panel) and $(k_{F} a_{F})^{-1} = 0.25$ (lower panel).}
\label{fig-II-14}
\end{center}
\end{figure}

A direct comparison of the temperature dependence of $\Delta_{\mathrm{large}}$ and  $\Delta_{\mathrm{pg}(-)}$ is shown in Fig.~\ref{fig-II-14}
for the two couplings previously considered.
Here, squares represent the values of $\Delta_{\mathrm{large}}$ obtained from the fittings reported in Fig.~\ref{fig-II-13}, circles are the values of $\Delta_{\infty}$ obtained independently from the expression (\ref{delta-infinity}), and triangles are the values of $\Delta_{\mathrm{pg}(-)}$ determined from the fittings (\ref{dispersion-relations}) to the dispersions.

It is evident from this figure that $\Delta_{\mathrm{pg}(-)}$ is a much faster decreasing function of temperature than $\Delta_{\infty}$, which
reflects the slow decay of the contact $C$ at high temperature \cite{PPPS-2010}.
Note also that some discrepancy arises between the values of $\Delta_{\mathrm{large}}$ and $\Delta_{\infty}$ at increasing temperature. 
This is due to the fact that for increasing temperature the interval of $k$ from which $\Delta_{\infty}$ can confidently be extracted should be centered progressively at a larger value of $k$, while in Fig.~\ref{fig-II-14} we have kept it at $2 k_{F} \le k \le 4 k_{F}$ for all temperatures.
In any case, the difference between $\Delta_{\mathrm{large}}$ and $\Delta_{\infty}$ is significantly smaller than that between $\Delta_{\infty}$ and
$\Delta_{\mathrm{pg}(-)}$.
Note finally that at unitarity $\Delta_{\mathrm{pg}(-)}$ and $\Delta_{\infty}$ almost coincide with each other close to $T_{c}$.
In this case, a single value can be effectively associated with the two energy scales.

However, the two energy scales soon deviate from each other not only for increasing temperature above $T_{c}$, but also away from unitarity on the two sides of the crossover.
For instance, in the BCS regime $\Delta_{\infty} = 2 \pi|a_{F}| n / m$ for $T \lesssim (m a_{F}^{2})^{-1}$ while the pseudo-gap would be exponentially small in the coupling parameter $(k_{F} a_{F})^{-1}$.
In the BEC regime, on the other hand, $\Delta_{\infty} = \sqrt{4 \pi n / (m^{2} a_{F})}$ for $T \lesssim (m a_{F}^{2})^{-1}$ while in this case the
``real'' gap in the single-particle excitations would equal half the value of the binding energy $(m a_{F}^{2})^{-1}$ of a composite boson
\cite{PPS-2009}.
 
It should be mentioned in this context that, by the alternative t-matrix approach of Ref.\cite{Levin-2010}, a trace of the pair-fluctuation 
propagator (quite similar to Eq.(\ref{delta-infinity}) for $\Delta_{\infty}^{2}$) was interpreted as representing (the square of) a pseudo-gap energy for all couplings and temperatures above $T_{c}$, thus making in practice the high-energy scale $\Delta_{\infty}$ and the low-energy pseudo-gap $\Delta_{\mathrm pg}$ to coincide with each other. 
This marks a difference between the approach of Ref.\cite{Levin-2010} and the present one, which keeps instead the two energy scales $\Delta_{\infty}$ and $\Delta_{\mathrm pg}$ distinct from each other.
 
\section{III. The repulsive case: Crossing vs avoided crossing}

We pass now to consider the occurrence of the two energy scales $\Delta_{\mathrm{pg}}$ and $\Delta_{\infty}$ from a different perspective, which emphasizes the differences one finds near $k_{F}$ for the two branches $\omega_{(\pm)}(k)$ when considering a Fermi gas with attractive \emph{or} repulsive inter-particle interaction.
These differences are related to the presence near $k_{F}$ of a finite \emph{or} vanishing value of $\Delta_{\mathrm{pg}}$, while the behavior of $\omega_{(\pm)}(k)$ for $k \gg k_{F}$ which is related to $\Delta_{\infty}$ remains essentially the same in the two cases.

Accordingly, we shall contrast the behavior near $k_{F}$ that will result from the single-particle spectral function $A(k,\omega)$ for the repulsive case, with that identified already in the previous Section for the attractive case.
In this way, we shall significantly extend the discussion given in Ref.\cite{Randeria-2010} on $A(k,\omega)$ for the repulsive case, that was there considered only for large values $k \gg k_{F}$. 

\vspace{0.05cm}
\begin{center}
{\bf A. Specular comparisons}
\end{center}
\vspace{0.05cm}

To make a meaningful comparison, we consider the attractive case with $a_{F} < 0$ and the repulsive case with $a_{F} > 0$ for \emph{the same value} $(k_{F} |a_{F}|)^{-1}$ of the dimensionless coupling. 
In the Appendix we provide the necessary analytic details for the less familiar repulsive case, at the level of the t-matrix approximation that we use in the numerical calculations.

As we have already discussed, one can not only determine the dispersions of the two peaks of $A(k,\omega)$ but also keep track of their weights.
In the attractive case considered in Section~II, this combined information has resulted in the phenomenon of \emph{avoided crossing} which typically occurs when two quantum levels with the same symmetry evolve as a function of a parameter \cite{LL-QM}.
In this case, pairing fluctuations induce above $T_{c}$ the same kind of particle-hole mixing which is characteristic of the BCS theory below $T_{c}$ \cite{Schrieffer-1964}. 
This mixing, in turn, makes the two branches of $A(k,\omega)$ to share the same symmetry (being partially particle-like and partially hole-like), in such a way that no crossing of the two dispersions occurs (the region of their minimum approach being, by definition, associated with the pseudo-gap).

The phenomenon of avoided crossing has to be contrasted with what happens instead in a Fermi liquid, a system where particle and hole excitations do not mix.
In this case, the two branches of $A(k,\omega)$ are expected to cross each other at $k_{F}$, where they abruptly exchange their weights in a similar fashion to what occurs typically for a \emph{crossing} \cite{LL-QM}.

As emphasized in Ref.\cite{Randeria-2010}, when $k \gg k_{F}$ the occurrence of two branches in $A(k,\omega)$ in the place of a single one even for a Fermi liquid stems from the requirement that for $k \gg k_{F}$ the wave-vector distribution $n(k)$ at zero temperature has a tail $\propto C/k^{4}$, in accordance with Tan's argument \cite{Tan-2008-I,Tan-2008-II}.
However, the branch at negative $\omega$ has an extremely small (albeit non-vanishing) weight, as we shall explicitly verify for a Fermi gas with a short-range repulsion.

\vspace{0.2cm}
\vspace{0.05cm}
\begin{center}
{\bf B. Working procedures}
\end{center}
\vspace{0.05cm}

It is discussed in the Appendix that an appropriate choice of the parameters $k_{0}/k_{F}$, $(k_{F} a_{F})^{-1}$, and $m k_{F} v_{0}$ entering 
Eq.(\ref{regularization}) has to be made for the repulsive case.

In particular, when exploring the region $k \gg k_{F}$ in order to extract the quantity $\Delta_{\infty}$, the values of $k$ should not be smaller than, say, $4 k_{F}$ (a value consistent with the plots reported in Fig.3 of the second of Refs.\cite{PPPS-2010}, where the contact $C$ was calculated numerically within the t-matrix approximation).
If we choose $m k_{F} v_{0}$ not larger than $10$ for speeding up the summations over the Matsubara frequencies, from Eq.(\ref{solution-regularization}) of the Appendix we obtain $(k_{F} a_{F})^{-1} \simeq 4.4$ to be an optimal value of the coupling to the purpose.
These values for the parameters have also been used in Fig.~\ref{figA3} of the Appendix, where the area of the peak of 
$A(k,\omega)$ at negative energies yields the value $\Delta_{\infty}/E_{F} = 0.1102$.
As expected in this rather extreme weak-coupling regime, a corresponding calculation made for the attractive case with coupling $(k_{F} a_{F})^{-1} = - 4.4$ yields the comparable value $\Delta_{\infty}/E_{F}= 0.0860$ [cf. Eq.(\ref{Delta-infinity-from-Galitskii})].

On the other hand, when exploring the region $k \approx k_{F}$ to focus on the issue raised above about ``crossing vs avoided crossing'', in the repulsive case the coupling $(k_{F} a_{F})^{-1}$ should not exceed, say, the value $1.5$.
Otherwise, in the corresponding calculation for the attractive case it would be extremely difficult to detect numerically the occurrence of an avoided crossing when $(k_{F} a_{F})^{-1}$ becomes smaller than $-1.5$. 
When $(k_{F} a_{F})^{-1}=1.5$, the value of $m k_{F} v_{0}$ has to be pushed up to $20$ to get $k_{0}/k_{F}(=1.37)$ larger than $k_{F}$ as required.

\vspace{0.05cm}
\begin{center}
{\bf C. Results for dispersions, weights, and widths}
\end{center}
\vspace{0.05cm}

Figure~\ref{fig-III-1} compares dispersions, weights, and widths of the two structures of $A(k,\omega)$ close to $k_{F}$, as obtained in the repulsive case with $(k_{F}a_{F})^{-1} = 1.5$ and $T=0$ (left panels) and in the attractive case with $(k_{F}a_{F})^{-1} = -1.5$ and $T=T_{c}=0.0576 T_{F}$ (right panels).
This comparison highlights and contrasts the essential characteristics found in $A(k,\omega)$ for a Fermi liquid with no pseudo-gap (left panels) and for a non-Fermi liquid with a pseudo-gap (right panels), at the corresponding values of $(k_{F}|a_{F}|)^{-1}$.
Note in particular that:

\begin{figure}[h]
\begin{center}
\includegraphics[angle=0,width=8.8cm]{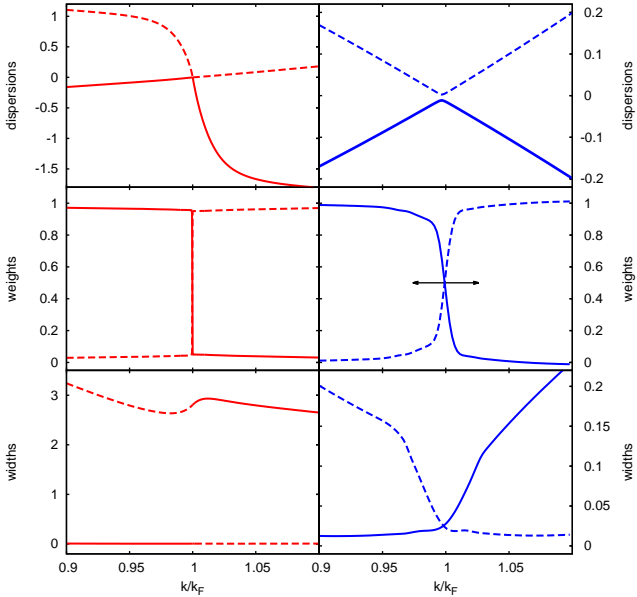}
\caption{Dispersions, weights, and widths extracted from the two features of $A(k,\omega)$ near $k_{F}$ for a Fermi system with: 
$(k_{F}a_{F})^{-1} = 1.5$, $m k_{F} v_{0}=20$, $k_{0} = 1.37 k_{F}$, and $T=0$ (left panels);
$(k_{F}a_{F})^{-1} = -1.5$ and $T=T_{c}=0.0576 T_{F}$ (right panels).
Energies are in units of $E_{F}$.
Full (dashed) lines correspond to the structure of $A(k,\omega)$ at $\omega < 0$ ($\omega > 0$).
The meaning of the double arrow is explained in the text.}
\label{fig-III-1}
\end{center}
\end{figure}

\noindent
(i) The crossing at $k_{F}$ of the dispersions of the two structures of $A(k,\omega)$ in the repulsive case (left-upper panel) contrasts with the avoided crossing in the attractive case (right-upper panel), whereby the two branches exchange their role and remain separated by the amount $2 \Delta_{\mathrm{pg}}$ (in the present case, the wave vector $k_{L}$ at which the avoided crossing occurs between the two branches in the attractive case is quite close to $k_{F}$);
\vspace{0.1cm}

\noindent
(ii) The behavior of the spectral weights associated with the two structures of $A(k,\omega)$ (for $\omega < 0$ and $\omega > 0$, respectively) shows an abrupt exchange at $k_{F}$ in the repulsive case (left-middle panel) which is typical of a level crossing, while a smooth evolution over a spread $\delta k$ such that $\delta k^{2}/(2 m) \approx \Delta_{\mathrm{pg}}$ results 
in the attractive case (the size of $\delta k$ is represented by a double arrow in the right-middle panel). 
[In the present case, $\Delta_{\mathrm{pg}}/E_{F} = 0.012$.]
We have further verified that the size of the abrupt jump $Z (\simeq 0.89)$ of the weight at $k_{F}$ in the repulsive case coincides (within $5 \%$) with the value obtained from Fermi liquid theory \cite{Fermi-liquids-books}, which is related to the (retarded) self-energy according to the expression $Z^{-1} = 1 - \left[ \partial \mathrm{Re}\{\Sigma(k,\omega)\} / \partial \omega \right]_{k=k_{F},\omega=0}$;
\vspace{0.1cm}

\noindent
(iii) The large differences between the widths of the two structures of $A(k,\omega)$ in the repulsive case which persist across 
$k_{F}$ (left-lower panel) strongly deviate from the attractive case (right-lower panel), where the widths of the two branches reflect into each other at $k_{F}$.

It should be remarked that the coupling $(k_{F}a_{F})^{-1} = -1.5$ we have used to obtain the right panels of Fig.~\ref{fig-III-1} is somewhat extreme, because very close to $k_{F}$ the sum of the widths of the two peaks of $A(k,\omega)$ exceeds their separation and the two peaks merge in a single one.
To continue discerning two separate peaks in this narrow range of wave vectors near $k_{F}$ (specifically, from $k \simeq 0.95 k_{F}$ up to $k \simeq 1.5 k_{F}$), a two-Lorentzian fit to the single broad peak is required.

\begin{figure}[t]
\begin{center}
\includegraphics[angle=0,width=6.0cm]{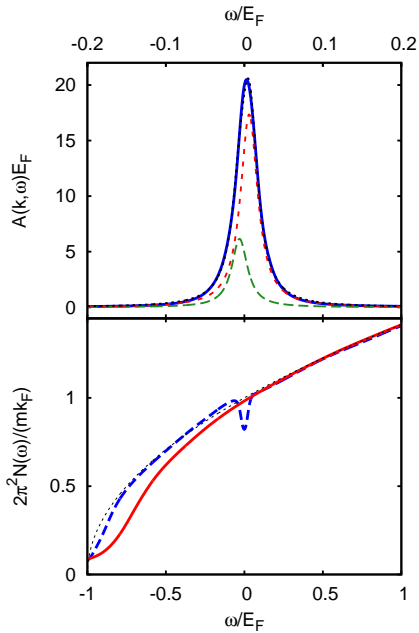}
\caption{Upper panel: Two-Lorentzian fit (short- and long-dashed lines) of the single broad feature of $A(k=k_{F},\omega)$ (full line) in the attractive case with $(k_{F}a_{F})^{-1} = -1.5$ and $T=T_{c}$.
Lower panel: Density of states vs $\omega$ for an attractive Fermi gas at $T=T_{c}$ with $(k_{F}a_{F})^{-1} = -1.5$ (dashed line),
and for a repulsive Fermi gas at $T=0$ with $(k_{F}a_{F})^{-1} = 1.5$ (full line).
The values of $v_{0}$ and $k_{0}$ are the same of Fig.~\ref{fig-III-1}.
The dotted line shows the result for a free Fermi gas.}
\label{fig-III-2}
\end{center}
\end{figure}

An example of this fit is shown at $k_{F}$ in the upper panel of Fig.~\ref{fig-III-2}, where the peak at higher energy (short-dashed line) has larger weight than the peak at lower energy (long-dashed line) since $k_{F} > k_{L}$ (consistently with the right panels of Fig.~\ref{fig-III-1}).
Typical values of the $\chi^{2}$-fit to isolate the two Lorentzians do not exceed $10^{-5}$.
It is through this kind of fit that we were able to identify the value of the pseudo-gap $\Delta_{\mathrm{pg}}$ reported above, even in this rather extreme situation.

\begin{figure}[h]
\begin{center}
\includegraphics[angle=0,width=5.5cm]{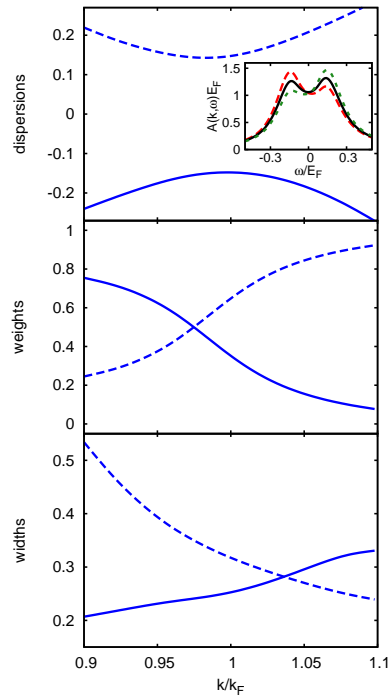}
\caption{Dispersions, weights, and widths at $T=T_{c}$ extracted from the two peaks of $A(k,\omega)$ about $k_{L}/k_{F} = 0.997 \pm 0.001$ in the attractive case with $(k_{F}a_{F})^{-1} = -0.8$.
Energies are in units of $E_{F}$.
The inset shows the profiles of $A(k,\omega)$ vs $\omega$ for three wave vectors $k / k_{F}= (0.980,0.992,1.004)$ that correspond to dashed, full, and dotted lines, in the order.}
\label{fig-III-3}
\end{center}
\end{figure}

Although in the attractive case near $k_{F}$ the two peaks of $A(k,\omega)$ merge apparently into a single one as a consequence of their broadening, the single-particle density of states $N(\omega)$ given by Eq.(\ref{DOS}) still maintains a well-pronounced feature about $\omega = 0$ due to the underlying pseudo-gap. 
This is evidenced by the dip occurring in the dashed line of the lower panel of Fig.~\ref{fig-III-2} corresponding to the attractive case, whose width is about $0.1 E_{F}$.
By contrast, the dip is absent in the full line in the lower panel of Fig.~\ref{fig-III-2}, which corresponds to the repulsive case and reproduces near 
$\omega = 0$ the free-fermion result per spin component (dotted line) whereby $A(\mathbf{k},\omega) = \delta(\omega - \xi_{\mathbf{k}})$.

The identification of the pseudo-gap is more direct when considering values less extreme than $(k_{F}a_{F})^{-1} = -1.5$ for the attractive coupling, for which two distinct peaks in $A(k,\omega)$ can be distinguished even near $k_{L}$.
This is shown in Fig.~\ref{fig-III-3} near $k_{F}$ for the coupling $(k_{F}a_{F})^{-1} = -0.8$ still on the BCS side of the crossover, for which we get $\Delta_{\mathrm{pg}}/E_{F} =0.127 \pm 0.005$.
As already mentioned, even when the two peaks can be clearly distinguished in the region of the avoided crossing, a two-Lorentzian fit improves the accuracy of the determination of the dispersions, weights, and widths associated with these peaks. 
Note from Fig.~\ref{fig-III-3} that, already at this less extreme coupling, moderate deviations from a simple BCS form arise in the dispersions and weights.

\section{IV. Concluding remarks}
\label{sec:conclusions}

In this paper, we have focused a great deal of our attention on the ``pseudo-gap physics'' which results near $k_{F}$ from the effects of attractive pairing fluctuations above $T_{c}$.
From a careful analysis of the single-particle spectral function $A(k,\omega)$, we have identified the essential characteristics of $A(k,\omega)$ that can meaningfully be transposed above $T_{c}$ starting from a mean-field description below $T_{c}$, even in situations when the inter-particle coupling is quite strong.
Accordingly, the occurrence of a pseudo-gap has enabled us to carry over above $T_{c}$ in an approximate way concepts and results that are firmly established below $T_{c}$, where a truly long-range order extends over the entire system.

We have further contrasted the occurrence of the pseudo-gap near $k_{F}$, which produces a characteristic back-bending of the dispersion relation associated with the peak of $A(k,\omega)$ at negative $\omega$, with the eventual evolution of this peak for $k \gg k_{F}$.
Along these lines, we have regarded the pseudo-gap energy $\Delta_{\mathrm{pg}}$ as being associated with pair correlations established in the system over intermediate distances of the order of the inter-particle spacing $k_{F}^{-1}$, while the contact physics for $k \gg k_{F}$ results from pair correlations of shorter range.
We have thus shown that attractive and repulsive pair correlations have similar effects for $k \gg k_{F}$ but yield drastically different results near 
$k_{F}$.

To appreciate intuitively on physical grounds the occurrence of pseudo-gap phenomena established by attractive pairing fluctuations above the critical temperature $T_{c}$ of the superfluid transition, it may be useful to draw an analogy with the occurrence of damped spin waves which are present in ferromagnetic materials above the Curie temperature $T_{C}$ when a strict long-range order is absent.
As it is often the case when dealing with phase transitions, examples from magnetic transitions may help one envisaging related phenomena occurring in different kinds of transitions.

In the case of a ferromagnet, what is lost above $T_{C}$ is the long-range order which organizes the spins over the whole sample.
Yet, the spins can remain organized over moderate distances, so that spin-wave excitations may still propagate over these distances. 
The lack of full long-range order manifests itself in the damping of these ``local'' spin waves, which broadens their frequency spectrum.
Experimentally, it has been long known since the work of Ref.\cite{Mook-1973} that spin waves can survive in magnetic materials above the transition temperature 
(see also the more recent work of Ref.\cite{Demmel-2007} and the references quoted therein).  
The accepted physical explanation for this effect is that a local magnetic order may exists above the magnetic transition temperature, such that spin waves can be supported over these length scales although with a shortened lifetime (this is especially true in low-dimensional systems - cf. Ref.\cite{MSS-1993}).  
Quite interestingly, experimental evidence has recently been collected that short-range spin waves may even underlie the mechanism for 
high-temperature superconductivity \cite{Miller-2011}. 

With this analogy in mind, pseudo-gap phenomena in a Fermi system with a mutual attractive inter-particle interaction can be envisaged as due to the persistence of a ``local pairing order'' above the superfluid temperature $T_{c}$, which takes place even when the (off-diagonal) long-range order is absent.
The local order which is preserved by pairing fluctuations above $T_{c}$ makes the single-particle excitation spectrum to resemble the one below $T_{c}$, although with an appreciable frequency broadening due to the temporal decay of these local excitations which cannot propagate over long distances in the absence of long-range order.

In addition, this analogy may help one appreciating the connection between the presence of a pseudo-gap at low energy and the contact $C$ at high energy \cite{PPPS-2010}, since they both derive from the existence of spatial correlations between fermions with opposite spins over moderate or else short distances.

It is, finally, relevant to recall that the persistence of spin waves in the normal phase of magnetic materials had spurred a strong debate in the literature, their evidence being disputed in favor of a broadening of the energy distribution \cite{Shirane-1983}.
Nevertheless, the existence of spin waves in the normal phase was eventually accepted by the occurrence of new direct experimental evidence \cite{Demmel-2007}.

Something similar is apparently going on at present for the occurrence of a pseudo-gap in the normal phase of a Fermi gas with attractive inter-particle interaction, its evidence from momentum-resolved photoemission experiments 
\cite{Jin-2008,JILA-Cam-2010,Cam-JILA-2011} having been disputed in favor of a more conventional Fermi-liquid picture from thermodynamic measurements \cite{Salomon-2010}.
Similarly to what happened for spin waves above $T_{C}$ in magnetic materials, the current dispute about pseudo-gap phenomena above $T_{c}$ could possibly be brought to an end by a new evidence coming from direct detection of the \emph{upper branch} 
of the dispersion of $A(k,\omega)$.
This should be possible for a Fermi gas with attractive interaction via the experimental technique of Ref.\cite{Jin-2008} once the temperature is raised sufficiently to populate the upper branch \cite{JILA-Cam-2010}, or for condensed-matter systems where hole-like bands may give similar access to unoccupied state.
 
\vspace{0.2cm}

This work was partially supported by the Italian MIUR under Contract Cofin-2007 ``Ultracold Atoms and Novel Quantum Phases''.
                                                                                                                                                                                                                                                                                                                                                                                                          
\appendix
\section{APPENDIX: T-MATRIX APPROXIMATION FOR A DILUTE FERMI GAS WITH REPULSIVE INTERACTION}

For a repulsive interaction of strength $v_{0} >0 $, the cutoff $k_{0}$ in Eq.(\ref{regularization}) cannot be let $\rightarrow \infty$ 
in order to have a (positive) non-vanishing scattering length $a_{F}$.
For this reason, a potential of finite range ($\approx k_{0}^{-1}$) has to be retained, without reaching the limit of a truly contact potential.
In this case, one expects quite generally $a_{F}$ to be small, of the order of $k_{0}^{-1}$ \cite{Fano-Rau}.
Solving for $k_{0}$ in Eq.(\ref{regularization}) one obtains:
\begin{equation}
k_{0} \, = \, \frac{\pi}{2 \, a_{F}} \, - \, \frac{2 \, \pi^{2}}{m \, v_{0}} \,\, .                                 \label{solution-regularization}
\end{equation}

\noindent
This relation sets the upper bound $\pi/(2 k_{F} a_{F})$ on $k_{0}/k_{F}$, which is reached for very large values of $m k_{F} v_{0}$.
In practice, we are interested in determining the behavior of the structures of the single-particle spectral function $A(k,\omega)$ for $k \approx k_{F}$ or larger, in such a way that $k_{0}$ should exceed these values of $k$.
On the other hand, too large values of $m k_{F} v_{0}$ (exceeding, say, 20) make the numerical summation over Matsubara frequency in Eq.(\ref{Matsubara-self-energy}) exceedingly difficult.
As a consequence, a compromise has to be reached about the values of the coupling $(k_{F} a_{F})^{-1}$ and $m k_{F} v_{0}$.

A systematic way to deal with finite values of the cutoff $k_{0}$ and the positive interaction strength $v_{0}$ is to introduce a \emph{separable} potential in $k$-space, of the form:
\begin{equation}
V(\mathbf{k},\mathbf{k'}) \, = \, v_{0} \, \theta(k_{0} - |\mathbf{k}|) \, \theta(k_{0} - |\mathbf{k'}|)           \label{separable-potential}
\end{equation}

\noindent
$\theta(k)$ being the Heaviside step function.
In this case, the particle-particle ladder reads:
\begin{equation}
\Gamma_{0}(\mathbf{k},\mathbf{k'};\mathbf{q},\Omega_{\nu}) = \theta(k_{0} - |\mathbf{k}|) \, \theta(k_{0} - |\mathbf{k'}|) \,\,   
                                                                                \tilde{\Gamma}_{0}(\mathbf{q},\Omega_{\nu})              \label{particle-particle-repulsive}
\end{equation}

\noindent
where
\begin{eqnarray}
- \, \tilde{\Gamma}_{0}(\mathbf{q},\Omega_{\nu})^{-1} & = &  \frac{1}{v_{0}} \, + \, \int \! \frac{d\mathbf{k}}{(2 \pi)^{3}} \, \theta(k_{0} - |\mathbf{k}|)       \label{separable-particle-particle-ladder}  \\
& \times & k_{B} T  \, \sum_{n} \, G_{0}(\mathbf{k},\omega_{n}) \, G_{0}(\mathbf{q - k},\Omega_{\nu} - \omega_{n})                                                      \nonumber                                                                    
\end{eqnarray}

\noindent
takes place of the expression (\ref{particle-particle-ladder}) that holds in the limit $k_{0} \rightarrow \infty$.

Finite values of $k_{0}$ and $v_{0}$ affect also the asymptotic behavior of $\tilde{\Gamma}_{0}(\mathbf{q},\Omega_{\nu})$ for large $|\mathbf{q}|$ and $|\Omega_{\nu}|$, which is
now given by:
\begin{equation}
\tilde{\Gamma}_{0}(\mathbf{q},\Omega_{\nu}) \, \simeq \, - \, v_{0} \, + \, \frac{\alpha(\mathbf{q})}{\xi_{\mathbf{q}} - i \Omega_{\nu}}                 \label{asymptotic-particle-particle}
\end{equation}

\noindent
with
\begin{equation}
\alpha(\mathbf{q}) \, = \, v_{0}^{2} \int \! \frac{d\mathbf{k}}{(2 \pi)^{3}} \, \theta(k_{0} - |\mathbf{k}|) \, \left[ 1 \, - \, f(\xi_{\mathbf{k}}) \, - \, f(\xi_{\mathbf{q-k}})  \right]  . \label{alpha}
\end{equation}

\noindent
The self-energy 
\begin{eqnarray}
&& \Sigma(\mathbf{k},\omega_{n}) = -  \, \theta(k_{0} - |\mathbf{k}|) \int \! \frac{d\mathbf{q}}{(2 \pi)^{3}} \, k_{B} T  
                                                   \sum_{\nu} \, \tilde{\Gamma}_{0}(\mathbf{q},\Omega_{\nu})                                   \nonumber \\
&& \times \,\, e^{i \Omega_{\nu}\eta} \, G_{0}(\mathbf{q - k},\Omega_{\nu} - \omega_{n}) \, \left[2 \, - \, \theta(k_{0} - |\mathbf{q-k}|) \right]        \label{separable-Matsubara-self-energy}     
\end{eqnarray} 

\noindent
(with the distinct Hartree and Fock contributions) can then be conveniently organized as follows:
\begin{equation}
\Sigma(\mathbf{k},\omega_{n}) \, = \, \theta(k_{0} - |\mathbf{k}|) \, \left[ \Sigma_{0} \, + \, \Sigma'(\mathbf{k},\omega_{n}) \right] \,\, .         \label{decomposition-self-energy}
\end{equation}

\noindent
Here, $\Sigma_{0} = v_{0} n_{f} /2$ with $n_{f}$ given by Eq.(\ref{density-zeroth-level}) in the zero-temperature limit of interest and for $k_{0} > \sqrt{2 m \mu}$, while 
\begin{eqnarray}
&&\Sigma'(\mathbf{k},\omega_{n}) \, = \, -  \, \int \! \frac{d\mathbf{q}}{(2 \pi)^{3}} \, k_{B} T  \, \sum_{\nu} \, \left[ \tilde{\Gamma}_{0}(\mathbf{q},\Omega_{\nu}) \, + \, v_{0} \right]    \nonumber  \\
&& \times \,\,\, G_{0}(\mathbf{q - k},\Omega_{\nu} - \omega_{n}) \, \left[2 \, - \, \theta(k_{0} - |\mathbf{q-k}|) \right]            \label{Sigma-prime}
\end{eqnarray}

\noindent
where the convergence factor $e^{i \Omega_{\nu}\eta}$ has been dropped owing to the $\Omega_{\nu}^{2}$-decay of the summand for large 
$|\Omega_{\nu}|$.

We are once more interested in the limit when $k^{2}/(2m)$ or $|\omega_{n}|$ are much larger than $k_{B} T$ and $\mu$.
In this limit, the self-energy (\ref{decomposition-self-energy}) reduces to the form:
\begin{equation}
\Sigma(\mathbf{k},\omega_{n}) \, \simeq \, -  \, \theta(k_{0} - |\mathbf{k}|) 
\left[ \frac{1}{2} n_{f} \tilde{\Gamma}_{0}(\mathbf{k},\omega_{n}) + \Delta_{\infty}^{2} \, G_{0}(\mathbf{k},-\omega_{n}) \right]         \label{asymptotic-decomposition-self-energy}
\end{equation}

\noindent
in analogy to Eq.(\ref{approximate-Matsubara-self-energy}), where $\Delta_{\infty}^{2}$ is still given by Eq.(\ref{delta-infinity}) with $ \tilde{\Gamma}_{0}(\mathbf{q},\Omega_{\nu})$
replacing  $\Gamma_{0}(\mathbf{q},\Omega_{\nu})$.

To obtain the single-particle spectral function according to Eq.(\ref{single-particle-spectral-function}), analytic continuation $i \omega_{n} \rightarrow \omega + i \eta$ to the real frequency axis needs to be performed.
The imaginary part of the retarded self-energy is now given by:
\begin{eqnarray}
&& \mathrm{Im}\Sigma(\mathbf{k},\omega) \, = \, -  \, \theta(k_{0} - |\mathbf{k}|) \int \! \frac{d\mathbf{q}}{(2 \pi)^{3}}  \left[2 \, - \, \theta(k_{0} - |\mathbf{q}|) \right]     \nonumber  \\
&& \times \,\,\, \mathrm{Im} \tilde{\Gamma}_{0}(\mathbf{k+q},\omega + \xi_{\mathbf{q}}) \, \left[ f(\xi_{\mathbf{q}}) + b(\omega + \xi_{\mathbf{q}}) \right]            \label{imaginary-separable-self-energy}
\end{eqnarray} 

\noindent
where $b(E) = (e^{E/(k_{B}T)} - 1)^{-1}$ is the Bose function, in terms of which the real part $\mathrm{Re}\Sigma(\mathbf{k},\omega)$ can be obtained via Kramers-Kronig transform.
To obtain the above expression we have used the spectral representation
\begin{equation}
\tilde{\Gamma}_{0}(\mathbf{q},\Omega_{\nu}) \, = \, - \, v_{0} \, - \,  \int_{-\infty}^{+\infty} \! \frac{d \omega}{\pi} \, 
\frac{\mathrm{Im} \tilde{\Gamma}_{0}(\mathbf{q},\omega)}{i \Omega_{\nu} \, - \, \omega} \,\, ,                                      \label{spectral-representation-Gamma}
\end{equation}

\noindent
which generalizes to the present context an analogous expression considered in Ref.\cite{PPSC-2002} for an attractive zero-range potential.

\begin{figure}[h]
\begin{center}
\includegraphics[angle=0,width=5.0cm]{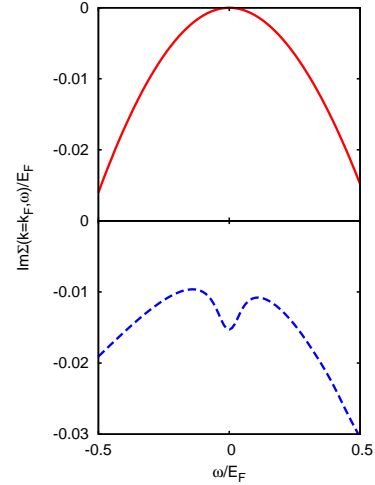}
\caption{Imaginary part of the retarded self-energy vs $\omega$ at $k=k_{F}$ for a Fermi gas with repulsive (upper panel) and attractive (lower panel) interaction.
Upper panel: $T=0$, $(k_{F}a_{F})^{-1} = 1.5$, $m k_{F} v_{0}=20$, and $k_{0} = 1.37 k_{F}$.
Lower panel: $T=T_{c}=0.0576 T_{F}$ and $(k_{F}a_{F})^{-1} = -1.5$.}
\label{figA1}
\end{center}
\end{figure}

It is interesting to note in this context that $f(\xi_{\mathbf{q}}) + b(\xi_{\mathbf{q}}) = 0$ in the zero-temperature limit, so that from Eq.(\ref{imaginary-separable-self-energy}) one obtains $\mathrm{Im}\Sigma(\mathbf{k},\omega = 0) = 0$ in that limit.
The vanishing of $\mathrm{Im}\Sigma(\mathbf{k},\omega = 0)$ is characteristic of a Fermi liquid at zero temperature for which a repulsive interaction is appropriate, and distinguishes it from a Fermi system with attractive interaction where superfluidity sets in at the critical temperature $T_{c}$.
A comparison of $\mathrm{Im}\Sigma(\mathbf{k},\omega)$ over an extended range of $\omega$ about $\omega = 0$, for a repulsive interaction at $T=0$ and an attractive interaction at $T_{c}$, is shown in Fig.~\ref{figA1} when $k=|\mathbf{k}|=k_{F}$ for the characteristic couplings $(k_{F}a_{F})^{-1} = \pm1.5$, 
in the order.

\begin{figure}[t]
\begin{center}
\includegraphics[angle=0,width=9.0cm]{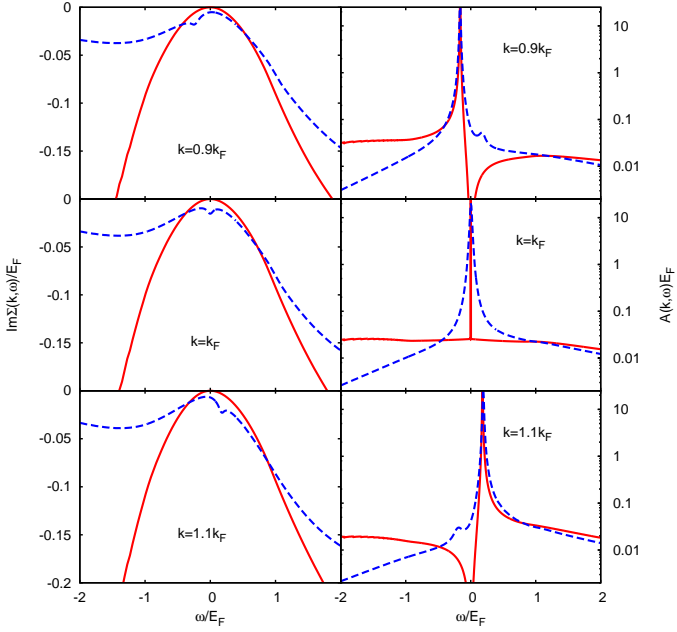}
\caption{Imaginary part of the retarded self-energy (left panels) and single-particle spectral function (right panels) vs $\omega$ for three values 
of $k$, corresponding to a Fermi gas with repulsive (full lines) and attractive (dashed lines) interaction with the same parameters of Fig.~\ref{figA1}.
The log scale in the right panels highlights the small weights of the tails.}
\label{figA2}
\end{center}
\end{figure}

These differences in $\mathrm{Im}\Sigma(\mathbf{k},\omega)$ between the repulsive and attractive cases result in marked differences in the single-particle spectral function $A(\mathbf{k},\omega)$, as shown in Fig.~\ref{figA2} over an even more extended range of $\omega$ for three distinct values of $k$ about $k_{F}$ and with the same parameters of Fig.~\ref{figA1}.

A comment about the structure of $A(\mathbf{k},\omega)$ at negative $\omega$ for \emph{large} $k$ is in order at this point.
Quite generally, $\mathrm{Im} \tilde{\Gamma}_{0}(\mathbf{k+q},\omega + \xi_{\mathbf{q}})$ in the expression (\ref{imaginary-separable-self-energy}) is non-vanishing provided the
condition
\begin{equation}
\omega + \xi_{\mathbf{q}} \ge \frac{(\mathbf{k + q})^{2}}{4 m} \, - \, 2 \mu                   \label{condition}
\end{equation}

\noindent
is satisfied.
In addition, when $\omega < 0$ the sum of the Fermi and Bose functions in Eq.(\ref{imaginary-separable-self-energy}) requires that $\mu < \frac{\mathbf{q}^{2}}{2m} < \mu - \omega$.
When $\omega < 0$, both conditions imply that $\frac{(\mathbf{k + q})^{2}}{4m} - 2 \mu < 0$.
For large $|\mathbf{k}|$, this means that $|\mathbf{q}|$ should be comparable with $|\mathbf{k}|$, and therefore that $\mathrm{Im}\Sigma(\mathbf{k},\omega)$ of
Eq.(\ref{imaginary-separable-self-energy}) is non-vanishing only for $|\omega| \gapprox \frac{(\mathbf{k})^{2}}{2m} - \mu$.
A more stringent condition on the frequency interval where $\mathrm{Im}\Sigma(\mathbf{k},\omega) \ne 0$ (and thus $A(\mathbf{k},\omega) \ne 0$) for $\omega < 0$ and large $|\mathbf{k}|$ was considered in Ref.\cite{Randeria-2010}.

Some additional technical differences, between our approach and the treatment of Ref.\cite{Randeria-2010} on the large-$k$ behavior of the single-particle spectral function for a Fermi gas 
with repulsive interaction, need also be mentioned.
In that work, emphasis was given to the large-$k$ behavior of $A(k,\omega)$ in the repulsive case, to show that a structure at negative $\omega$ develops also in
this case whose spectral weight is proportional to the contact $C$.
For that purpose, it was sufficient to consider coupling values $(k_{F} a_{F})^{-1} \gapprox + 10$ corresponding to an extremely diluted situation, for which the details of the inter-particle potential of short range are irrelevant.
In that limit, it was possible in Ref.\cite{Randeria-2010} to use for the particle-particle ladder an expansion of the expression (\ref{particle-particle-ladder}) up to second order in $a_{F}$ that would hold, in principle, only in the attractive case. 
For these rather extreme coupling values, the results obtained in Ref.\cite{Randeria-2010} coincide with ours, which are obtained through a separable potential with parameters
$k_{0}$ and $v_{0}$ related via Eq.(\ref{solution-regularization}).
We recall that our use of a separable potential is required by the fact that we have extended the calculation of $A(k,\omega)$ in the repulsive case down to much smaller values ($\approx  +1$) of the coupling $(k_{F} a_{F})^{-1}$, in order to be able to compare with the corresponding results in the attractive case when $k \approx k_{F}$.

\begin{figure}[t]
\begin{center}
\includegraphics[angle=0,width=7.5cm]{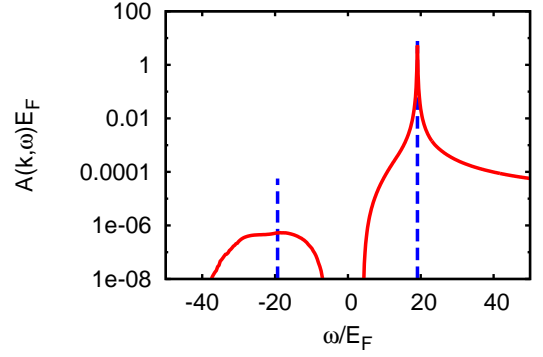}
\caption{Single-particle spectral function of a Fermi gas with repulsive interaction $(k_{F}a_{F})^{-1} = 4.4$ at $T = 0$ for $k = 4.5 k_{F}$ (here, 
$k_{0} = 5 k_{F}$ and $m k_{F} v_{0}=10$). The broad peaks from the complete calculations (full lines) are contrasted with the delta-like spikes (dashed lines) resulting when in the self-energy  the large-$k$ limit is taken before analytic continuation to real frequencies.}
\label{figA3}
\end{center}
\end{figure}

As a final comment, we warn that, by performing the analytic continuation to the real frequency axis directly on the asymptotic form (\ref{asymptotic-decomposition-self-energy}) of the self-energy for large $k < k_{0}$, one would get an expression of the type (\ref{approximate-spectral-density}) for the single-particle spectral function with a sharp peak at $\omega = - \xi_{\mathbf{k}}$.
The correct procedure, of first performing the analytic continuation on the self-energy (\ref{Sigma-prime}) and only then taking its limit for large 
$k$, results instead in a broad structure of the single-particle spectral function at negative frequencies, although with the same area 
$\Delta_{\infty}^{2}/(4 \xi_{\mathbf{k}}^{2})$ of the approximate form (\ref{approximate-spectral-density}) for the attractive case.
This is explicitly shown in Fig.~\ref{figA3} for $k = 4.5 k_{F}$ when $(k_{F}a_{F})^{-1} = 4.4$.
This kind of non-commutativity, between taking the analytic continuation and performing the large-$|\mathbf{k}|$ expansion on the self-energy, was already pointed out in Ref.\cite{PPS-2004} while extending the t-matrix approach to the superfluid phase below $T_{c}$.

Although for large $k$ the feature in $A(k,\omega)$ at negative $\omega$ has formally the same weight $\Delta_{\infty}^{2}/(4 \xi_{\mathbf{k}}^{2})$ in the attractive and repulsive cases, the numerical value of $\Delta_{\infty}$ (and thus of the contact $C$) depends on the sign of the interaction and differs in the two cases.
For instance, we obtain $\Delta_{\infty}/E_{F} = 0.0860$ for the attractive case and $\Delta_{\infty}/E_{F} = 0.1102$ for the repulsive case using the parameters of Fig.~\ref{figA1}.
These values can be compared with those obtained analytically from the Galitskii theory of a dilute Fermi gas \cite{Galitskii}, according to which to  leading orders in $k_{F} a_{F}$:
\begin{equation}
\Delta_{\infty} \simeq \frac{2 \pi n |a_{F}|}{m} \, \left[ 1 + \, \frac{6}{35 \pi} (11 - 2 \ln 2) k_{F} a_{F} \right]        
                                                                                                                                                        \label{Delta-infinity-from-Galitskii}
\end{equation}

\noindent
where only the second term within braces depends on the sign of $a_{F}$.
This expression yields $\Delta_{\infty}/E_{F} = 0.0835$ for the attractive case and $\Delta_{\infty}/E_{F} = 0.1063$ for the repulsive case, values which compare well with those obtained above from our numerical calculations.




\begin{thebibliography}{99}

\bibitem{Jin-2008} J. T. Stewart, J. P. Gaebler, and D. S. Jin, Nature {\bf 454}, 744 (2008).

\bibitem{JILA-Cam-2010} J. P. Gaebler, J. T. Stewart, T. E. Drake, D. S. Jin, A. Perali, P. Pieri, and G. C. Strinati, 
                                         Nature Phys. {\bf 6}, 569 (2010). 
                                         
\bibitem{Levin-2005}  A common ground of the pseudo-gap physics in the context of the BCS-BEC crossover, from 
                                   high-temperature superconductors to ultra-cold Fermi gases, has been emphasized by Q. Chen, 
                                   J. Stajic, S. Tan, and K. Levin, Phys. Rep. {\bf 412}, 1 (2005).                                         

\bibitem{Tan-2008-I} S. Tan, Ann. Phys. {\bf 323}, 2952 (2008).

\bibitem{Tan-2008-II} S. Tan, Ann. Phys. {\bf 323}, 2971 (2008).

\bibitem{PPS-2009} P. Pieri, A. Perali, and G. C. Strinati, Nature Phys. {\bf 5}, 736 (2009).

\bibitem{PPSC-2002} A. Perali, P. Pieri, G. C. Strinati, and C. Castellani, Phys. Rev. B {\bf 66}, 024510 (2002).

\bibitem{Cam-JILA-2011} A. Perali, F. Palestini, P. Pieri, G. C. Strinati, J. T. Stewart, J. P. Gaebler, T. E. Drake, 
                                         and D. S. Jin, Phys. Rev. Lett. {\bf 106}, 060402 (2011).
                                         
\bibitem{Schrieffer-1964} J. R. Schrieffer, \emph{Theory of Superconductivity} (Benjiamin, New York, 1964).                                          

\bibitem{BCS-50-years} See also, L. N. Cooper and D. Feldman, Eds., \emph{BCS: 50 Years} (World Scientific, Singapore, 2011).

\bibitem{Randeria-2010} W. Schneider and M. Randeria, Phys. Rev. A {\bf 81}, 021601(R) (2010).

\bibitem{Levin-2004} J. Stajic, J. N. Milstein, Q. Chen, M. L. Chiofalo, M. J. Holland, and K. Levin, Phys. Rev. A {\bf 69}, 063610 (2004).

\bibitem{PPPS-2004} A. Perali, P. Pieri, L. Pisani, and G. C. Strinati, Phys. Rev. Lett. {\bf 92}, 220404 (2004).

\bibitem{Micnas-1990} R. Micnas, J. Ranninger, and S. Robaszkiewicz, Rev. Mod. Phys. {\bf 62}, 113 (1990).

\bibitem{Randeria-1992} M. Randeria, N. Trivedi, A. Moreo, and R. T. Scalettar, Phys. Rev. Lett. {\bf 69}, 2001 (1992).

\bibitem{Levin-1997} B. Jank\'{o}, J. Maly, and K. Levin, Phys. Rev. B {\bf 56}, 11407(R) (1997).

\bibitem{Bulgac-2009} P. Magierski, G. Wlazlowski, A. Bulgac, and J. E. Drut, Phys. Rev. Lett. {\bf 103}, 210403 (2009).

\bibitem{Ohashi-2009} S. Tsuchiya, R. Watanabe, and Y. Ohashi, Phys. Rev. A {\bf 80}, 033613 (2009).

\bibitem{Levin-2010} C.-C. Chien, H. Guo, Y. He, and K. Levin, Phys. Rev. A {\bf 81}, 023622 (2010).

\bibitem{Sheehy-2010} S.-Q. Su, D. E. Sheehy, J. Moreno, and M. Jarrell, Phys. Rev. A {\bf 81}, 051604(R) (2010).

\bibitem{Ohashi-2010} S. Tsuchiya, R. Watanabe, and Y. Ohashi, Phys. Rev. A {\bf 82}, 033629 (2010).

\bibitem{Bulgac-2011} P. Magierski, G. Wlazlowski, and A. Bulgac, Phys. Rev. Lett. {\bf 107}, 145304 (2011).

\bibitem{FW} See, e.g., A. L. Fetter and J. D. Walecka, \emph{Quantum Theory of Many-Particle Systems} 
                     (McGraw-Hill, New York, 1971), Sect. 49.
                      
\bibitem{PPS-2004} P. Pieri, L. Pisani, and G. C. Strinati, Phys. Rev. B {\bf 70}, 094508 (2004).                                           
                      
\bibitem{GPS-2008} See, e.g., S. Giorgini, L. P. Pitaevskii, and S. Stringari, Rev. Mod. Phys. {\bf 80}, 1215 (2008).      
                                                                    
\bibitem{footnote-1} Quite generally, one may add to the right-hand side of Eq.(\ref{dispersion-relations}) a constant shift
                                $\omega_{(\pm)}^{(0)}$ for the two branches, with the consequence that different values of $k_{L (\pm)}$     
                                will result from the new fittings. These shifts are certainly appropriate when comparing the calculated dispersions with 
                                the experimental data, as it was done in Ref.\cite{Cam-JILA-2011}, but appears here superfluous since the 
                                emphasis is about the (approximate) validity of the BCS-like expression (\ref{dispersion-relations}) internally to the 
                                calculated spectra.
                                
\bibitem{Fermi-liquids-books} P. Nozi\`eres, \emph{Theory of Interacting Fermi Systems} (Addison-Wesley, Reading, MA, 1964);
                                               E. M. Lifshitz and L. P. Pitaevskii, \emph{Statistical Physics: Theory of the Condensed State} 
                                               (Butterworth-Heinemann, Oxford, 1980).                            

\bibitem{Randeria-1993} C. A. R. S\'{a} de Melo, M. Randeria, and J. R. Engelbrecht, Phys. Rev. Lett. {\bf 71}, 3202 (1993). 

\bibitem{footnote-3} For large negative $\omega$, it results from Eqs.(\ref{DOS}) and (\ref{approximate-spectral-density}) that 
                                $N(\omega) \cong (2m)^{3/2} \Delta_{\infty}^{2} / (16 \pi^{2} |\omega|^{3/2})$.

\bibitem{PPPS-2010} Z. Yu, G. M. Bruun, and G. Baym, Phys. Rev. A {\bf 80}, 023615 (2009);
                                   F. Palestini, A. Perali, P. Pieri, and G. C. Strinati, Phys. Rev. A {\bf 82}, 021605(R) (2010); 
                                   E. D. Kuhnle, S. Hoinka, P. Dyke, H. Hu, P. Hannaford, and C. J. Vale, Phys. Rev. Lett. {\bf 106}, 170402 (2011);Ê
                                   H. Hu, X.-J. Liu, Êand P. D. Drummond, New J. Phys. {\bf 13}, 035007 (2011);
                                   J. E. Drut, T.  A. L\"{a}hde, and T. Ten, Phys. Rev. Lett. {\bf 106}, 205302 (2011).

\bibitem{LL-QM} See, e.g., L. D. Landau and L. M. Lifshitz, \emph{Quantum Mechanics (Non-Relativistic Theory)} 
                           (Butterworth-Heinemann, Amsterdam, 2003), Sect. 79.    

\bibitem{Mook-1973} H. A. Mook, J. W. Lynn, and R. M. Nicklow, Phys. Rev. Lett. {\bf 30}, 556 (1973).

\bibitem{Demmel-2007} F. Demmel and T. Chatterji, Phys. Rev. B {\bf 76}, 212402 (2007).

\bibitem{MSS-1993} N. Majlis, S. Selzer, and G. C. Strinati, Phys. Rev. B {\bf 48}, 957 (1993).

\bibitem{Miller-2011} J. Miller, Phys. Today, September 2011, p. 13, and references therein. 

\bibitem{Shirane-1983} O. Steinsvoll, C. F. Majkrzak, G. Shirane, and J. Wicksted, Phys. Rev. Lett. {\bf 51}, 300 (1983). 

\bibitem{Salomon-2010} S. Nascimb\`ene, N. Navon, K. J. Jiang, F. Chevy, and C. Salomon, Nature {\bf 463}, 1057 (2010).

\bibitem{Fano-Rau} See, e.g., U. Fano and A. R. P. Rau, \emph{Atomic Collisions and Spectra} (Academic Press, New York, 1986), Chap. 4.
                                
\bibitem{Galitskii} V. M. Galitskii, Sov. Phys. JETP {\bf 7}, 104 (1958).

\end{thebibliography}
\end{document}